\documentclass[aps,showpacs,nofootinbib,amsmath,amssymb,preprint]{revtex4}
\usepackage{bm}
\usepackage{graphicx}
\usepackage{amsthm}

\usepackage{color}

\begin{document}

\title{
Backreaction of Mass and Angular Momentum Accretion on Black Holes: 
General Formulation of the Metric Perturbations and 
Application to the Blandford-Znajek Process
}

\author{
Masashi Kimura${}^{1}$,
Tomohiro Harada${}^{1}$,
Atsushi Naruko${}^{2}$
and 
Kenji Toma${}^{3,4}$
}

\affiliation{
${}^{1}$ Department of Physics, Rikkyo University, Tokyo 171-8501, Japan
\\
${}^{2}$ Center for Gravitational Physics, Yukawa Institute for Theoretical Physics, Kyoto University, Kyoto 606-8502, Japan
\\
${}^{3}$ Frontier Research Institute for Interdisciplinary Sciences, Tohoku University, Sendai 980-8578, Japan
\\
${}^{4}$ Astronomical Institute, Graduate School of Science, Tohoku University, Sendai 980-8578, Japan
}

\date{\today}
\preprint{RUP-21-7}
\preprint{YITP-21-43}

\begin{abstract}
We study the metric backreaction of mass and angular momentum accretion on black holes.
We first develop the formalism of 
monopole and dipole linear gravitational perturbations around the Schwarzschild black holes in the Eddington-Finkelstein coordinates
against the generic time-dependent matters.
We derive the relation between
the time dependence of the mass and angular momentum of the black hole and the energy-momentum tensors of accreting matters.
As a concrete example, we apply our formalism to the Blandford-Znajek process 
 around the slowly rotating black holes.
We find that the time dependence of  the monopole and dipole perturbations 
can be interpreted as the slowly rotating Kerr metric with time-dependent mass and spin parameters,
which are determined from the energy and angular momentum 
extraction rates of the Blandford-Znajek process.
We also show that the Komar angular momentum and the area of the apparent horizon are decreasing and increasing in time, respectively, while they are consistent with the Blandford-Znajek argument of energy extraction in term of black hole mechanics 
if we regard the time-dependent mass parameter as the energy of the black hole.
\end{abstract}

\maketitle

\section{Introduction}
Black holes in astrophysical situations are usually assumed to be the Kerr black holes,
and the matter fields are treated as test fields.
This is because the effects of matter distribution on the curvature are typically small,
and then the spacetime is determined from the vacuum Einstein equations
which only admit the Kerr black holes as stationary regular black holes
due to the uniqueness theorem in general relativity~\cite{Israel:1967wq, Robinson:1975bv, Carter:1971zc}.
Nevertheless, if we take into account the effect of matter distribution on the spacetime,
we can discuss the effect of energy-momentum tensor on the metric 
by the gravitational perturbations around the background black holes.
In particular, if matter accretions on black holes exist, we expect that 
the mass and angular momentum of black holes secularly change.
In this paper, we would like to make clear this issue by explicitly 
studying the gravitational perturbations around black holes.
As a first step, we focus on the case of the Schwarzschild black hole background.

The linear gravitational perturbations around the Schwarzschild black holes 
were studied by Regge and Wheeler~\cite{Regge:1957td}, and Zerilli~\cite{Zerilli:1970se, Zerilli:1971wd}.
For the higher-order multipole perturbations, 
where the degrees of freedoms of the gravitational waves exist,
the linearized Einstein equations reduce to the second order wave equations called the Regge and Wheeler, and 
Zerilli equations with the source terms~\cite{Regge:1957td, Zerilli:1970se, Zerilli:1971wd, Martel:2005ir}.
Because now we are interested in the evolution of the black hole mass and angular momentum by
the matter accretion, we need to study the monopole and dipole perturbations.
In~\cite{Dokuchaev:2011gt, Babichev:2012sg},
the monopole perturbations against the generic stationary accreting matters around the Schwarzschild black holes were studied.
Recently, in~\cite{Nakamura:2021mfv}, it was shown that 
both monopole and dipole perturbations for the generic time-dependent matters
around the Schwarzschild black holes
can be solved in a static coordinate system.
In this paper, we extend the formalism in~\cite{Babichev:2012sg}, where the Eddington-Finkelstein coordinates are used, to the case of monopole and dipole perturbations for the generic time-dependent accreting matters.
To study the evolution of the mass and angular momentum of black holes, 
the regularity of the accreting matters on the black hole horizon is required
and the Eddington-Finkelstein coordinates are suitable for checking the regularity.
As shown in Sec.~\ref{sec:schperturbation}, we derive the relation between
the time dependence of the mass and angular momentum of the black hole and the energy-momentum tensors.

As interesting phenomena around the rotating black holes, we can consider the energy extraction from black holes,
not only increasing the mass of the black holes.
The energy extraction by test particles is known as the Penrose process~\cite{Penrose:1969pc, Wagh:1986tsa},
and that by the scattering waves is the superradiance~\cite{Zeldovich:1971, Zeldovich:1972, Starobinsky:1973aij, Starobinsky:1974, Lasota:2013kia, Brito:2015oca}. 
The energy extraction process by the force-free electromagnetic fields is 
the Blandford-Znajek process~\cite{Blandford:1977ds}, which is 
a candidate of the central engine for gamma-ray bursts and active galactic nucleus jets.
The various aspects of the Blandford-Znajek process were studied in~\cite{McKinney:2004ka, Jacobson:2017xam, Komissarov:2008yh, Koide:2014xpa, Tanabe:2008wm, Pan:2015iaa, Armas:2020mio, Toma:2014kva, Toma:2016jmz, Kinoshita:2017mio, Noda:2019mzd}.
In this paper, we
discuss the metric backreaction of the energy extraction from rotating black holes
by the Blandford-Znajek process.
Because the discussions by the Blandford and Znajek in~\cite{Blandford:1977ds} are based 
on the slow rotation approximation of the Kerr black holes,
we discuss the backreaction using the non-linear gravitational perturbations around the Schwarzschild black holes,
where both the effects of the slow rotation and the backreaction of the Blandford-Znajek process are taken into account.
In the study of the non-linear gravitational perturbations, at each order,
we need to solve equations whose forms are same as those of linear order 
but the non-linear effects appear in the source terms. For this reason, our formalism can be applied to this problem.

This paper is organized as follows. 
In Sec.~\ref{sec:schperturbation}, we develop the formalism by extending the discussion in~\cite{Babichev:2012sg}.
In Sec.~\ref{sec:bz}, we briefly review the force-free electromagnetic fields considered by the Blandford and Znajek~\cite{Blandford:1977ds}.
Applying the formalism in Sec.~\ref{sec:schperturbation} to 
the electromagnetic fields in Sec.~\ref{sec:bz},
we study the metric backreaction of the Blandford-Znajek process in Sec.~\ref{sec:backreaction}.
The black hole mechanics in this situation is discussed in~\ref{sec:bhmechanics}.
Sec.~\ref{sec:summary} presents summary and discussions.
We use the units in which $c = G = 1$.

\section{The Backreaction of the Mass and Angular Momentum Accretion on Schwarzschild Black Holes}
\label{sec:schperturbation}
Let us consider the situation where the effect of matter distribution on curvature is weak. 
Then, we need to solve the Einstein equations
\begin{align}
G_{\mu \nu} =  8\pi \epsilon T_{\mu \nu},
\end{align}
with the small parameter $\epsilon$. 
At the lowest order ${\cal O}(\epsilon^0)$, the metic is given by a vacuum solution of the Einstein equations.
For the later convenience, in this section, we choose the lowest-order vacuum solution 
as the Schwarzschild metric $g_{\mu \nu}^{\rm Sch}$
\begin{align}
g_{\mu \nu}^{\rm Sch}dx^\mu dx^\nu = -f dt^2
+f^{-1}dr^2 + r^2(d\theta^2 + \sin^2\theta d\phi^2),
\end{align}
with $f = 1- r_0/r$ and $r_0 = 2M$, where $M$ denotes the background black hole mass.
When we consider the effect of $\epsilon T_{\mu \nu}$, 
the spacetime will be described by the metric with the small deviation from the Schwarzschild metric
\begin{align}
g_{\mu \nu} = g_{\mu \nu}^{\rm Sch} + \epsilon h_{\mu \nu}.
\label{fullmetric1}
\end{align}
The Einstein tensor of this metric becomes
\begin{align}
G_{\mu \nu} &= \epsilon \delta G_{\mu \nu} 
\notag\\&
= \epsilon \left[
-\frac{1}{2}\nabla_\mu \nabla_\nu h^\alpha{}_\alpha
-
\frac{1}{2}\nabla^\alpha \nabla_\alpha h_{\mu \nu}
+
\nabla^\alpha \nabla_{(\mu}h_{\nu)\alpha}
+
\frac{1}{2}g_{\mu \nu}(
\nabla^\alpha \nabla_\alpha h^\beta{}_\beta
-
\nabla^\alpha \nabla^\beta h_{\alpha \beta})
\right]
+
{\cal O}(\epsilon^2).
\notag\\&
=: \epsilon {\cal L}^{\rm Sch}[h_{\alpha \beta}]_{\mu \nu}
+
{\cal O}(\epsilon^2),
\label{linearEinsteineq}
\end{align}
where $\nabla_\mu$ denotes the the covariant derivative of the Schwarzschild metric $g_{\mu \nu}^{\rm Sch}$,
and we raise or lower indices by $g_{\mu \nu}^{\rm Sch}$.
At the order of ${\cal O}(\epsilon)$, we need to solve the equations
\begin{align}
\epsilon {\cal L}^{\rm Sch}[h_{\alpha \beta}]_{\mu \nu} = 8\pi \epsilon  T_{\mu \nu}.
\label{perturbationeqsch}
\end{align}
The energy-momentum tensor satisfies
\begin{align}
\nabla^\mu T_{\mu \nu} = 0,
\end{align}
due to the Bianchi identity.
We should note that the following discussion holds if only the basic equations formally take the form of Eq.~\eqref{perturbationeqsch}. In particular, when we discuss the effect of the backreaction for the Blandford-Znajek process in Sec.~\ref{sec:backreaction},
we will solve Eq.~\eqref{perturbationeqsch} with the effective energy-momentum tensor.

For the spherically symmetric spacetime background, we can decompose tensor quantities by 
the tensor spherical harmonics characterized by $\ell, m$ ($\ell = 0,1,2\cdots, m =0,\pm1,\cdots \pm\ell$), and we can separately discuss even and odd parities and different $\ell, m$ modes when we solve Eq.~\eqref{perturbationeqsch}~\cite{Regge:1957td, Zerilli:1970se, Zerilli:1971wd}.
In this section, we study $\ell = 0$ and $\ell = 1$ odd-parity time-dependent gravitational perturbations for generic time-dependent matter distribution because those modes are important for the study of the backreaction of 
accreting matters on the mass and angular momentum of black holes. 
In~\cite{Babichev:2012sg}, the case of stationary energy-momentum tensor was discussed, and
recently, in~\cite{Nakamura:2021mfv}, the generic time-dependent case was discussed in the static coordinate system.
In this paper, we work in the Eddington-Finkelstein coordinates $(V,r,\theta,\Phi)$ with $dV = dt + f^{-1}dr, d\Phi = d\phi$, 
and the line element becomes
\begin{align}
g_{\mu \nu}^{\rm Sch}dx^\mu dx^\nu = -f dV^2
+2 dVdr+ r^2(d\theta^2 + \sin^2\theta d\Phi^2).
\end{align}
In this coordinate system, it is easy to discuss the regularity of tensor quantities at $r = r_0$
because the finiteness of the tensor components at $r = r_0$ coincides with the regularity condition at the horizon.

\subsection{monopole perturbations}
The perturbed metric for $\ell = 0$ is given by
\begin{align}
h_{\mu \nu}^{(+)}|_{\ell=0} dx^\mu dx^\nu
&=
H_{0}(V,r)dV^2 + 2 H_{1}(V,r)dVdr, 
\label{perturbedmetric_lzero}
\end{align}
where we choose the gauge condition $h_{rr} = h_{\theta \theta} (=h_{\Phi \Phi}/\sin^2\theta) = 0$ (see Appendix.~\ref{appendix:gaugetr}).
In this gauge choice, there is a residual gauge mode $H_{0} \to H_{0} -2 f \eta(V), H_{1} \to H_{1} + \eta(V)$,
where $\eta(V)$ is an arbitrary function of $V$.
We note that this residual gauge mode corresponds to the rescale of the coordinate $V$.
The generic energy-momentum tensor for $\ell = 0$ becomes
\begin{align}
T_{\mu \nu}^{(+)}|_{\ell=0} dx^\mu dx^\nu
&=
T_{VV}(V,r)dV^2 + 2 T_{Vr}(V,r)dVdr + T_{rr}(V,r)dr^2 + T_{\Omega}(V,r) r^2(d\theta^2 + \sin^2\theta d\Phi^2).
\end{align}
The equation $\nabla^{\mu}T_{\mu V} = 0$ shows that the quantity
\begin{align}
{\cal A} = 4\pi r^2 (f T_{Vr} + T_{VV}),
\label{energyfluxA}
\end{align}
satisfies 
\begin{align}
\partial_r {\cal A} = -4\pi r^2\partial_V T_{Vr}.
\label{eq:dra}
\end{align}
The quantity ${\cal A}$ is interpreted as the accretion rate of the energy
\begin{align}
{\cal A} = \int_0^{2\pi} \int_0^{\pi} T_{\mu}{}^{\nu}(\partial_V)^\mu (dr)_{\nu} r^2\sin\theta d\theta d\Phi,
\end{align}
which is related to the flux associated with the conservation law $\nabla^\mu(T_{\mu \nu}(\partial_V)^\nu) = 0$.
We note that $(\partial_V)^\mu := \partial x^\mu/\partial V$ and $(dr)_{\nu}:= \partial r/\partial x^\mu$.
If the energy-momentum tensor is stationary, ${\cal A}$ becomes constant.
Introducing a quantity ${\cal E}$ as
\begin{align}
{\cal E} = \int_0^{2\pi} \int_0^{\pi} T_{\mu \nu}(\partial_V)^\mu (\partial_V)^\nu r^2\sin\theta d\theta d\Phi,
\end{align}
we can write Eq.~\eqref{eq:dra} as
\begin{align}
(f \partial_r + \partial_V) {\cal A} = \partial_V {\cal E}.
\label{eq:localenergyconservation}
\end{align}
In the static coordinates $(t,r,\theta,\phi)$, Eq.~\eqref{eq:localenergyconservation} becomes $f\partial_r {\cal A} = \partial_t {\cal E}$, then we can easily see that this corresponds to the local energy conservation law.\footnote{
In our definition, ${\cal A}$ is positive when a positive accretion into the black hole exists.
The equation can be written in the conventional conservation form $\partial_t {\cal E} + f\partial_r (-{\cal A}) = 0$.
}
The other components of the equations $\nabla^{\mu}T_{\mu \nu} = 0$ shows the relation among $T_{Vr}, T_{rr}$ and $T_{\Omega}$
\begin{align}
4 r T_{\Omega} - 2\partial_r(r^2 (T_{Vr} + f T_{rr})) - r^2 T_{rr} \partial_r f - 2 r^2 \partial_V T_{rr}= 0.
\end{align}
In the same manner as~\cite{Babichev:2012sg}, introducing new variables $\delta M(V,r)$ and $\lambda(V,r)$ as
\begin{align}
H_0(V,r) &= \frac{2 \delta M(V,r)}{r} + 2 f \lambda(V,r), 
\label{h0even}
\\
H_1(V,r) &= - \lambda(V,r),
\label{h1even}
\end{align}
the $(V,V), (V,r)$ and $(r,r)$ components of the Einstein equations give
\begin{align}
\partial_V \delta M &= {\cal A},
\\
\partial_r \delta M  &= -4\pi r^2 T_{Vr},
\\
\partial_r \lambda   &= - 4\pi r T_{rr}.
\end{align}
These equations can be solved as
\begin{align}
\delta M &= \delta m + \int_{V_0}^V {\cal A}(\bar{V},r) d\bar{V}  
- 4\pi \int_{r_0}^r \bar{r}^2 T_{Vr}(V_0,\bar{r}) d\bar{r},
\label{deltaMeven}
\\
\lambda &= -4\pi \int_{r_0}^r \bar{r} T_{rr}(V,\bar{r}) d\bar{r} + \chi(V),
\label{lambdaeven}
\end{align}
where $\delta m$ and $V_0$ are constants and $\chi(V)$ is an arbitrary function of $V$.
The function $\chi(V)$ corresponds to the residual gauge mode, {\it i.e.}, the rescaling of $V$. 
We note that the other components of the Einstein equations are automatically satisfied.
To summarize, the perturbed metric for $\ell =0$ is described by 
\begin{align}
h_{\mu \nu}^{(+)}|_{\ell=0} dx^\mu dx^\nu
=
\left(
 \frac{2 \delta M}{r} + 2 f \lambda
\right)dV^2 - 2 \lambda dVdr, 
\end{align}
where $\delta M$ and $\lambda$ are given by Eqs.~\eqref{deltaMeven} and \eqref{lambdaeven}.

If there do not exist $\ell \ge 1$ perturbations, due to the spherical symmetry of the spacetime, we can calculate the Misner-Sharp mass\footnote{
The Misner-Sharp mass for the spherically symmetric spacetime is given by $M_{\rm MS} = (1- |dr|^2)r/2$,
where $r$ is the area radius and $|dr|^2 = g^{\mu \nu}(dr)_\mu (dr)_\nu$.
} 
for the metric $g_{\mu \nu}^{\rm Sch} + \epsilon h_{\mu \nu}$
at  
$(V,r)$ as
\begin{align}
M_{\rm MS} &= M + \epsilon \delta M,
\end{align}
where $\delta M$ is given by Eq.~\eqref{deltaMeven}.
We can see that the constant $\delta m$ denotes the deviation of the Misner-Sharp mass from the background mass parameter $M$
at $V = V_0$ and $r = r_0$. 
We note that the degrees of freedom in choosing $\delta m$ and $V_0$ are degenerate because 
if we change $V_0$, $\delta m$ is shifted.
Also, the quantity ${\cal A}$ determines the time dependence of the mass
\begin{align}
\partial_V M_{\rm MS} = \epsilon {\cal A}.
\end{align}

\subsection{odd-parity dipole perturbations}
The perturbed metric for the $\ell = 1$ odd-parity modes is given by
\begin{align}
h_{\mu \nu}^{(-)}|_{\ell=1} dx^\mu dx^\nu
&=
4\sqrt{\pi/3}\, h_0(V,r) \sin\theta (\partial_{\theta}Y_{1,0}) 
dV d\Phi
\notag\\&=
-2  h_0(V,r) \sin^2\theta 
dV d\Phi,
\label{metricell1}
\end{align}
where $Y_{1,0} = 2^{-1}\sqrt{3/\pi}\cos \theta$,\footnote{The other cases $Y_{1,\pm1}$ can be obtained by acting the ladder operators of the spherical harmonics on the perturbed metric Eq.~\eqref{metricell1} if needed.}
and we choose
the gauge condition $h_{r\Phi} = 0$ (see Appendix.~\ref{appendix:gaugetr}).
In this gauge choice, there is a residual gauge mode, $h_0 \to h_0 +r^2 \zeta(V)$, where $\zeta(V)$
is an arbitrary function of $V$.
Note that this residual gauge mode corresponds to the 
shift of the coordinate $\Phi$ by the function of $V$.
The generic energy-momentum tensor for the $\ell = 1$ odd-parity modes becomes
\begin{align}
T_{\mu \nu}^{(-)}|_{\ell=1} dx^\mu dx^\nu
&=
-2 \sin^2\theta 
d\Phi[t_{V\Phi}(V,r)dV + t_{r\Phi}(V,r)dr].
\end{align}
The non-trivial component of $\nabla^{\mu}T_{\mu \nu} = 0$ shows that the quantity
\begin{align}
{\cal B} =  \frac{16\pi r^2}{3r_0} (t_{V\Phi} + f t_{r\Phi}),
\label{eq:defb}
\end{align}
satisfies 
\begin{align}
\partial_r {\cal B} = - \frac{16\pi r^2}{3r_0}\partial_V t_{r\Phi}.
\label{eq:drb}
\end{align}
The quantity ${\cal B}$ is interpreted as  
the accretion rate of the angular momentum
\begin{align}
{\cal B} =
- \frac{1}{M} \int_0^{2\pi} \int_0^{\pi} T_{\mu}{}^{\nu}(\partial_\Phi)^\mu (dr)_{\nu} r^2\sin\theta d\theta d\Phi,
\end{align}
where $M = r_0/2$, and $(\partial_\Phi)^\mu := \partial x^\mu/\partial \Phi$.
This is related to the flux associated with the conservation law $\nabla^\mu(T_{\mu \nu}(\partial_\Phi)^\nu) = 0$.
When the energy-momentum tensor is stationary, ${\cal B}$ becomes constant.
Introducing a quantity ${\cal J}$ as
\begin{align}
{\cal J} = - \frac{1}{M}\int_0^{2\pi} \int_0^{\pi} T_{\mu \nu}(\partial_V)^\mu (\partial_\Phi)^\nu r^2\sin\theta d\theta d\Phi,
\end{align}
we can write Eq.~\eqref{eq:drb} as
\begin{align}
(f \partial_r + \partial_V) {\cal B} = \partial_V {\cal J}.
\label{eq:localangularmomentumconservation}
\end{align}
In the static coordinates $(t,r,\theta,\phi)$, 
Eq.~\eqref{eq:localangularmomentumconservation} becomes $f\partial_r {\cal B} = \partial_t {\cal J}$, then we can easily see that this corresponds to the local angular momentum conservation law.\footnote{
In our definition, ${\cal B}$ is positive when a positive angular momentum accretion onto the black hole exists.
The equation can be written in the conventional conservation form $\partial_t {\cal J}+ f\partial_r (-{\cal B}) = 0$.}
The $(r,\Phi)$ component of the Einstein equations becomes
\begin{align}
\partial_r^2 h_0 - \frac{2}{r^2}h_0 = 16\pi t_{r\Phi}.
\end{align}
The general solutions of this equation are given by
\begin{align}
h_0(v,r) = \frac{C_1(V)}{r} + r^2 C_2(V) + h_0^{\rm IH}(V,r),
\end{align}
where $C_1$ and $C_2$ are arbitrary functions of $V$, and $h_0^{\rm IH}$ is an inhomogeneous solution
\begin{align}
h_0^{\rm IH}(V,r) = 16\pi r^2 \int_{r_0}^r \frac{1}{\bar{r}^4} 
\left[\int_{r_0}^{\bar{r}} \tilde{r}^2 t_{r\Phi}(V,\tilde{r}) d\tilde{r} \right] d\bar{r}.
\label{eq:ihsol}
\end{align}
The other components of the Einstein equations give
\begin{align}
\partial_V C_1 = r_0 {\cal B}|_{r = r_0}.
\end{align}
The general solution of this equation is given by
\begin{align}
C_1 =r_0 \delta a +r_0 \int_{V_0}^V {\cal B}(\bar{V},r_0) d\bar{V},
\end{align}
where $\delta a$ is a constant.
To summarize, the perturbed metric becomes
\begin{align}
 h_{\mu \nu}^{(-)}|_{\ell=1} dx^\mu dx^\nu
&=
- 
\frac{2 r_0 \sin^2\theta}{r} 
d\Phi dV\left[ \delta a 
+  \int_{V_0}^V {\cal B}(\bar{V},r_0) d\bar{V} + \frac{r}{r_0}( h_0^{\rm IH} + r^2 C_2(V))
\right],
\label{eq:h-l1}
\end{align}
where the function $C_2(V)$ corresponds to the residual gauge mode.

If there do not exist $m\neq 0$ perturbations, we can calculate the Komar angular momentum 
associated with the Killing vector $\partial_\Phi$ 
for the metric $g_{\mu \nu}^{\rm Sch} + \epsilon  h_{\mu \nu}$
at the radius $r$ as
\begin{align}
J_{\rm Komar}
= 
\epsilon M \left[ \delta a +  \int_{V_0}^V {\cal B}(\bar{V},r_0) d\bar{V}
+
\frac{r}{6M}\Big(2h_0^{\rm IH}
-
r\partial_r h_0^{\rm IH}\Big)
\right].
\end{align}
We note that $h_0^{\rm IH} = \partial_r h_0^{\rm IH} = 0$ at $r = r_0$.
The time dependence of $J_{\rm Komar}$ at the radius $r$ becomes
\begin{align}
\partial_V J_{\rm Komar} &= \epsilon M {\cal B}.
\end{align}
We can see that $\delta a$ corresponds to a constant shift in the Kerr parameter for slowly rotating cases,
and ${\cal B}$ determines the time dependence of the angular momentum at the radius $r$.

\subsection{Remarks}
\subsubsection{Uniqueness of the Kerr metric if $T_{\mu \nu} = 0$ in the exterior regions for $V \ge V_1$}
Let us consider that the energy-momentum tensor $T_{\mu \nu}$ in $r \ge r_0$ vanishes for $V \ge V_1 (> V_0)$.
This corresponds to the situation that the matter fields are electrically neutral and 
they completely fall into the black hole at $V = V_1$.
In that case, according to our formalism, the perturbed metric for $\ell =0$ and odd-parity $\ell = 1$ modes
becomes that for the slowly rotating Kerr black holes for $r \ge r_0$ and $V \ge V_1$
\begin{align}
& (g_{\mu \nu}^{\rm Sch} + \epsilon h_{\mu \nu}^{(+)}|_{\ell=0} + \epsilon h_{\mu \nu}^{(-)}|_{\ell=1} )dx^\mu dx^\nu
\notag\\&
= 
-\left[
1 - \frac{2 M_{\rm [final]}}{r}
\right]dV^2
+2 dVdr 
+ r^2 (d\theta^2+\sin^2\theta d\Phi^2)
-
\frac{4 M  a_{\rm [final]}\sin^2\theta}{r} d\Phi dV,
\label{eq:finalstateSch}
\end{align}
with
\begin{align}
M_{\rm [final]} &= M + \epsilon \delta m + \epsilon \int_{V_0}^{V_1} {\cal A}(V,r_0) dV, 
\\
a_{\rm [final]} &=  \epsilon \delta a +  \epsilon \int_{V_0}^{V_1} {\cal B}(V,r_0)dV. 
\end{align}
Note that 
we can evaluate $\delta M$ in Eq.~\eqref{deltaMeven} at $r = r_0$ for $V \ge V_1$
because of the relation $\partial_r \delta M = -4\pi r^2\partial_V T_{Vr} = 0$ for $V \ge V_1$.
Thus, the integrals of ${\cal A}$ and ${\cal B}$ at $r = r_0$ give the changes of 
the mass and the angular momentum of the black hole, respectively.

\subsubsection{Vaidya metric}
The Vaidya metric~\cite{Vaidya:1951zz} is the exact spherically symmetric 
solution with the radiating matter
\begin{align}
T_{\mu \nu}dx^\mu dx^\nu = \frac{d{\cal M}(V)/dV}{4\pi r^2}dV^2.
\label{eq:null_dust}
\end{align}
On the other hand, using our formalism with Eq.~(\ref{eq:null_dust}), we obtain
\begin{align}
(g_{\mu \nu}^{\rm Sch} + \epsilon h_{\mu \nu}^{(+)}|_{\ell=0} )dx^\mu dx^\nu
&= 
-\left(
1 - \frac{2(M + \epsilon {\cal M}(V))}{r}
\right)dV^2
+2 dVdr + r^2 (d\theta^2+\sin^2\theta d\Phi^2).
\label{eq:vaidya}
\end{align}
Thus, we can see that our linear perturbation formalism reproduces 
the exact 
Vaidya metric~\cite{Vaidya:1951zz}.\footnote{
The metric in Eq.~\eqref{eq:vaidya} takes the Kerr-Schild form~\cite{KerrSchild:1965, KerrSchild:2009}.
It is known that the Einstein tensor of the Kerr-Schild form metric is linear to the unknown function (see {\it e.g.,} \cite{Myers:1986un, Anabalon:2009kq}).
This is the reason why the linear perturbation analysis can derive the exact solution.
}

\subsubsection{The conservation laws and fluxes}
\label{subsec:conservationlaw}
The quantities
${\cal A}$ in Eq.~\eqref{eq:dra} and ${\cal B}$ in Eq.~\eqref{eq:drb} are 
related to the energy and angular momentum fluxes associated with the
conservation laws $\nabla^\mu(T_{\mu \nu}(\partial_V)^\nu) = 0$ and $\nabla^\mu(T_{\mu \nu}(\partial_\Phi)^\nu) = 0$, respectively.
We should note that this discussion holds if only the basic equations formally take the form of Eq.~\eqref{perturbationeqsch}.
In particular, as is discussed later, 
the equations whose forms are same as Eq.~\eqref{perturbationeqsch},
but $T_{\mu \nu}$ is replaced by the effective energy momentum tensors $T_{\mu \nu}^{\rm eff}$,
appear in the context of the non-linear perturbations around the Schwarzschild metric.
In that case, equations $\nabla^\mu(T_{\mu \nu}^{\rm eff}(\partial_V)^\nu) = 0$ and $\nabla^\mu(T_{\mu \nu}^{\rm eff}(\partial_\Phi)^\nu) = 0$
hold for the covariant derivative with respect to the Schwarzschild metric, and the corresponding global conservation laws exist.

\section{The energy-momentum tensor of the Blandford-Znajek process}
\label{sec:bz}

\subsection{The force-free electromagnetic fields around the Kerr spacetime}

We consider the test electromagnetic field $F_{\mu \nu} = \partial_\mu A_\nu - \partial_\nu A_\mu$
with the electric current density vector $j^\mu$ on the Kerr spacetime.
In this section, $g_{\mu \nu}$ denotes the Kerr metric, 
and $\nabla_\mu$ denotes the corresponding covariant derivative.
The metric of the Kerr spacetime in the Boyer-Lindquist coordinates $(t,r,\theta,\phi)$ is given by
\begin{align}
g_{\mu \nu}dx^\mu dx^\nu 
&=
- \frac{\Delta -a^2 \sin^2\theta}{\Sigma}dt^2
-
\frac{2 a \sin^2\theta (r^2 + a^2 - \Delta)}{\Sigma} dt d\phi
+
\frac{\Sigma}{\Delta}dr^2 + \Sigma d\theta^2
\notag\\&\quad +
\frac{(r^2+a^2)^2 - \Delta a^2 \sin^2\theta}{\Sigma}
\sin^2\theta d\phi^2,
\label{metricKerrBL}
\end{align}
where $\Sigma$ and $\Delta$ are 
\begin{align}
\Sigma &= r^2 + a^2 \cos^2\theta,
\\
\Delta &= r^2 + a^2 - 2 M r.
\end{align}
The constants $M$ and $a$ denote the mass and spin parameters.
The black hole horizon locates at $r = r_+ = M + \sqrt{M^2 - a^2}$.
The Maxwell equations on this spacetime are given by
\begin{align}
\nabla^\mu F_{\mu \nu} = 4\pi j_\nu.
\label{eq:maxwelleq}
\end{align}
We note that the equations
\begin{align}
\nabla_{[\mu}F_{\nu \rho]} = 0,
\label{eq:maxwellidentity}
\end{align}
are automatically satisfied from $F_{\mu \nu} = \partial_\mu A_\nu - \partial_\nu A_\mu$.
The energy-momentum tensor of the electromagnetic field
\begin{align}
T_{\mu \nu}^{\rm EM} = 
F_{\mu \alpha}F_{\nu}{}^\alpha - \frac{1}{4}g_{\mu \nu}
F_{\alpha \beta}F^{\alpha \beta},
\label{eq:tmunuem}
\end{align}
satisfies
\begin{align}
\nabla^\mu T_{\mu \nu}^{\rm EM} &= - 4\pi F_{\nu \mu}j^\mu.
\label{eq:maxwelltmunudiv}
\end{align}
If the right hand side of Eq.~\eqref{eq:maxwelltmunudiv}, {\it i.e.,} the 
Lorentz force term, is neglected, 
the force-free condition 
\begin{align}
F_{\nu \mu}j^\mu &= 0,
\label{eq:lorentzforceneglect}
\end{align}
is satisfied.
Then, the energy-momentum tensor of the electromagnetic field satisfies
\begin{align}
\nabla^\mu T_{\mu \nu}^{\rm EM} &= 0.
\label{eq:emeom}
\end{align}
We should note that under the condition $T^{\rm particle}_{\mu\nu} \ll T^{\rm EM}_{\mu\nu}$, where $T^{\rm particle}_{\mu\nu}$ is the particle energy density,
the total energy momentum conservation equations $\nabla^\mu(T^{\rm particle}_{\mu\nu} + T^{\rm EM}_{\mu\nu}) = 0$ 
reduce to Eq.~\eqref{eq:emeom}. 
This implies that $T^{\rm particle}_{\mu\nu} \ll T^{\rm EM}_{\mu\nu}$
is the sufficient condition for Eq.~\eqref{eq:lorentzforceneglect}.
To summarize, the force-free electromagnetic fields $F_{\mu \nu}$ can be obtained by solving Eq.~\eqref{eq:emeom} with Eq.~\eqref{eq:tmunuem},
and the electric current density vector $j^\mu$ can be calculated from Eq~\eqref{eq:maxwelleq}.

Because the Boyer-Lindquist coordinates do not cover the black hole horizon, 
the location $r = r_+$ becomes a coordinate singularity
and tensors have apparently singular behavior there.
In order to solve this problem, we introduce 
the Kerr-Schild coordinates $(T,r,\theta,\Phi)$ by $dT = dt + 2 M r dr/\Delta, d\Phi = d\phi + a dr/\Delta$.
The Kerr metric in the Kerr-Schild coordinates becomes
\begin{align}
& g_{\mu \nu}^{\rm KS}dx^\mu dx^\nu =
-\frac{\Delta - a^2 \sin^2\theta}{\Sigma}dT^2
+
\frac{4 M r}{\Sigma}dT dr
-
\frac{4 M r}{\Sigma} a \sin^2\theta dT d\Phi
+
\left(1 + \frac{2 M r}{\Sigma}\right)dr^2 
\notag\\& + \Sigma d\theta^2
+
\frac{(r^2 + a^2)^2 - \Delta a^2 \sin^2\theta}{\Sigma} \sin^2\theta d\Phi^2
-
2 a \sin^2\theta\left(1 + \frac{2 M r}{\Sigma}\right) dr d\Phi.
\end{align}

\subsection{The Blandford-Znajek solutions in the the Kerr-Schild coordinates}

In~\cite{Blandford:1977ds}, Blandford and Znajek studied the stationary and axisymmetric 
force-free electromagnetic fields around the slowly rotating Kerr metric,
and the energy and angular momentum extraction though the magnetic fields,
called the Blandford-Znajek process.
In this paper, we focus on the so-called split-monopole solution,
and the solution in the Kerr-Schild coordinates is given by~\cite{Blandford:1977ds, McKinney:2004ka}
\begin{align}
T_{\mu \nu}^{\rm BZ} = 
 \left(F_{\mu \alpha}F_{\nu}{}^\alpha - \frac{1}{4}g_{\mu \nu}^{\rm KS}
F_{\alpha \beta}F^{\alpha \beta}\right),
\end{align}
where the explicit forms of $F_{\mu \nu}$ are
\begin{align}
F_{Tr} &= \omega \partial_r A_\Phi,
\\
F_{T\theta} &= \omega \partial_\theta A_\Phi,
\\
F_{T\Phi} &= 0,
\\
F_{r\theta} &= \sqrt{|\det(g_{\mu \nu}^{\rm KS})|} B_\Phi,
\\
F_{r\Phi}&= \partial_r A_\Phi,
\\
F_{\theta \Phi}&= \partial_\theta A_\Phi,
\end{align}
and
\begin{align}
\omega &=  \omega_1 \frac{a}{M}  + {\cal O}(a^3),
\\
B_\Phi &=  B_{\Phi1} \frac{a}{M} + {\cal O}(a^3), 
\\
A_\Phi &= A_{\Phi 0} +  A_{\Phi 2}\left(\frac{a}{M}\right)^2 + {\cal O}(a^4).
\end{align}
The functions $A_{\Phi 0}, A_{\Phi 2}, \omega_1$ and $B_{\Phi 1}$ are given by
\begin{align}
A_{\Phi 0} &= - C \cos \theta,
\\
A_{\Phi 2} &= C {\cal F}(r) \cos\theta \sin^2\theta,
\\
\omega_1 &= \frac{1}{8M},
\\
B_{\Phi 1} &= \frac{-C}{8Mr^2} \left(1 + \frac{4M}{r}\right),
\end{align}
where $C$ is implicitly assumed to be of different signs for different signs of $\cos\theta$,
and the function ${\cal F}(r)$ is a regular solution of the differential equation
\begin{align}
\frac{d^2{\cal F}}{dr^2} + \frac{2M}{r(r-2M)} \frac{d{\cal F}}{dr}
-
\frac{6{\cal F}}{r(r-2M)}
+
\frac{M(r+2M)}{r^3(r-2M)} = 0.
\end{align}
The explicit form of ${\cal F}(r)$ is
\begin{align}
{\cal F}(r) &= \left[
{\rm Li}_2\left(\frac{2M}{r}\right) - \ln\left(1-\frac{2M}{r}\right)\ln \frac{r}{2M} 
\right]
\frac{r^2 (2r-3M)}{8M^3}
\notag\\&+
\frac{M^2+3 M r - 6r^2}{12M^2} \ln \frac{r}{2M}
+
\frac{11}{72}
+
\frac{M}{3r} + \frac{r}{2M} - \frac{r^2}{2M^2},
\end{align}
where ${\rm Li}_2$ is the second polylogarithm function
\begin{align}
{\rm Li}_2(x) = - \int_{0}^1 dt \frac{\ln(1-tx)}{t}.
\end{align}
The asymptotic behaviors of ${\cal F}$ at $r = 2M$ and $r = \infty$ are
\begin{align}
{\cal F} &= \frac{6\pi^2 - 49}{72} + \frac{6\pi^2 - 61}{24M}(r-2M) + {\cal O}((r-2M)^2),
\\
{\cal F} &= \frac{M}{4r} + \frac{M^2\ln (r/M)}{10 r^2} - \frac{M^2(11 + 20 \ln 2)}{200 r^2} + {\cal O}((\ln r)/r^3),
\end{align}
respectively.
We can confirm that $T_{\mu \nu}^{\rm BZ}$ satisfies the 
equations for the force-free electromagnetic fields~\eqref{eq:emeom}.\footnote{
We note that $F_{\mu \nu}$ satisfies Eq.~\eqref{eq:maxwellidentity},
and this implies that $A_\mu$ with $F_{\mu \nu} = \partial_\mu A_\nu - \partial_\nu A_\mu$ exists.}
Also $F_{\mu \nu}$ satisfies the degenerate condition 
\begin{align}
\star F^{\mu \nu}F_{\mu \nu}=0,
\label{eq:degenerate}
\end{align}
where
$\star F_{\mu \nu} = F^{\alpha \beta}\epsilon_{\alpha \beta \mu \nu}/2$ and $\epsilon_{\alpha \beta \mu \nu}$
is the Levi-Civita tensor.\footnote{The degenerate condition Eq.~\eqref{eq:degenerate} 
can be derived from the force-free condition Eq.~\eqref{eq:lorentzforceneglect} for non-zero $j^\mu$ (see {\it e.g.}~\cite{Kinoshita:2017mio}). We also note that Eq.~\eqref{eq:degenerate} 
is compatible with the ideal magnetohydrodynamic (MHD) condition~\cite{McKinney:2004ka}.
}
As shown in~\cite{Blandford:1977ds, McKinney:2004ka},
the energy and angular momentum extraction rate are given by
\begin{align}
 \dot{E}_{\rm BZ} &:= -  \int_0^{2\pi} \int_0^{\pi}\sqrt{|\det(g_{\mu \nu}^{\rm KS})|}\, T^{\rm BZ}_{\mu}{}^\nu 
(\partial_T)^\mu (dr)_\nu d\theta d\Phi 
= \frac{\pi}{24} \frac{a^2 C^2}{M^4} + {\cal O}(a^4).
\label{eq:energyexrateBZ}
\\
 \dot{J}_{\rm BZ} &:=   \int_0^{2\pi} \int_0^{\pi} \sqrt{|\det(g_{\mu \nu}^{\rm KS})|}\, T^{\rm BZ}_{\mu}{}^\nu 
(\partial_\Phi)^\mu (dr)_\nu
d\theta d\Phi
 = \frac{\pi}{3} \frac{a C^2}{M^2}+ {\cal O}(a^3).
\label{eq:amexrateBZ}
\end{align}
We can see that the relation $\dot{E}_{\rm BZ} = \omega \dot{J}_{\rm BZ}$ holds at this order.

\section{The Backreaction of the Blandford-Znajek Process}
\label{sec:backreaction}

\subsection{Perturbation scheme}

The discussions in Sec.~\ref{sec:bz} are based on the test field approximation around the Kerr black holes.
When we consider the backreaction of the Blandford-Znajek process on the spacetime,
we regard the parameter $C^2$ as a small parameter
so that the effect of the energy-momentum tensor $T^{\rm BZ}_{\mu \nu} (\propto C^2)$ on the spacetime is weak.
Introducing dimensionless small parameters $\alpha, \beta$ as\footnote{
The shape of the letter $\alpha$ is similar to $a$, 
and $\beta$ reminds us of the magnetic fields $B$.}
\begin{align}
\alpha &:= \frac{a}{M},
\\
\beta &:= \frac{C^2}{M^2},
\end{align}
the energy-momentum tensor $T_{\mu \nu}^{\rm BZ}$ can be written by the Taylor series around 
$(\alpha,\beta) = (0,0)$ as
\begin{align}
 T_{\mu \nu}^{\rm BZ} &=  \beta T_{\mu \nu}^{(0,1)} + \alpha \beta T_{\mu \nu}^{(1,1)} 
+  \alpha^2 \beta T_{\mu \nu}^{(2,1)} + {\cal O}(\alpha^3\beta).
\end{align}
To discuss the backreaction of the Blandford-Znajek process, we need to solve the Einstein equations
\begin{align}
G_{\mu \nu} = 8\pi T^{\rm BZ}_{\mu \nu} = 
 8\pi \beta  T_{\mu \nu}^{(0,1)} +  8\pi \alpha \beta  T_{\mu \nu}^{(1,1)} +  8\pi \alpha^2 \beta  T_{\mu \nu}^{(2,1)} 
+ {\cal O}(\alpha^3 \beta).
\end{align}
We expand the metric tensor as
\begin{align}
g_{\mu \nu} &= g_{\mu \nu}^{\rm Kerr} + g_{\mu \nu}^{\rm BZ}, 
\end{align}
with
\begin{align}
g_{\mu \nu}^{\rm Kerr} &= g_{\mu \nu}^{\rm Sch} + \alpha h_{\mu \nu}^{(1,0)} + \alpha^2 h_{\mu \nu}^{(2,0)} + {\cal O}(\alpha^3),
\label{eq:hn0}
\\
g_{\mu \nu}^{\rm BZ} &= \beta  h_{\mu \nu}^{(0,1)} + \alpha \beta  h_{\mu \nu}^{(1,1)} + \alpha^2 \beta  h_{\mu \nu}^{(2,1)} 
+ {\cal O}(\alpha^3 \beta).
\end{align}
The Einstein tensor becomes
\begin{align}
G_{\mu \nu} &= \beta  G_{\mu \nu}^{(0,1)} +  \alpha \beta  G_{\mu \nu}^{(1,1)} 
+ \alpha^2 \beta  G_{\mu \nu}^{(2,1)} 
+ {\cal O}(\alpha^3 \beta) + {\cal O}(\beta^2).
\end{align}
We note that the Einstein tensor at ${\cal O}(\beta^0)$ vanishes because
${\cal O}(\beta^0)$ metric is the Kerr metric.
At each order, we need to solve the following equations:
\begin{align}
\beta  G_{\mu \nu}^{(0,1)} &= 8\pi \beta T_{\mu \nu}^{(0,1)},
\\
\alpha \beta  G_{\mu \nu}^{(1,1)} &=8\pi \alpha \beta T_{\mu \nu}^{(1,1)},
\\
\alpha^2 \beta  G_{\mu \nu}^{(2,1)} &= 8\pi \alpha^2 \beta  T_{\mu \nu}^{(2,1)},
\end{align}
Schematically, we can write $G_{\mu \nu}^{(0,1)}, G_{\mu \nu}^{(1,1)}, G_{\mu \nu}^{(2,1)}$ as
\begin{align}
\beta G_{\mu \nu}^{(0,1)} &= \beta {\cal L}^{\rm Sch}[h^{(0,1)}_{\alpha \beta}]_{\mu \nu}
\\
\alpha \beta G_{\mu \nu}^{(1,1)} &= 
\alpha \beta {\cal L}^{\rm Sch}[h^{(1,1)}_{\alpha \beta}]_{\mu \nu} 
-
8\pi
\alpha \beta
\tilde{T}_{\mu \nu}^{(1,1)}
\\
\alpha^2 \beta G_{\mu \nu}^{(2,1)} &= \alpha^2 \beta {\cal L}^{\rm Sch}[h^{(2,1)}_{\alpha \beta}]_{\mu \nu} 
-
8\pi
\alpha^2 \beta
\tilde{T}_{\mu \nu}^{(2,1)}
\end{align}
where $\tilde{T}_{\mu \nu}^{(1,1)}$ and $\tilde{T}_{\mu \nu}^{(2,1)}$ denote the effects of the non-linear perturbation, {\it e.g.,} $\tilde{T}_{\mu \nu}^{(1,1)}$ is constructed from $h^{(1,0)}_{\mu \nu}$ and $h^{(0,1)}_{\mu \nu}$.
Thus, at each order, we solve the following equations which can be seen as {\it the 
linear perturbation around the Schwarzschild spacetime with the effective energy-momentum tensors}:
\begin{align}
\beta {\cal L}^{\rm Sch}[h^{(0,1)}_{\alpha \beta}]_{\mu \nu}  &=  8\pi \beta T_{\mu \nu}^{(0,1)}
\\
\alpha \beta {\cal L}^{\rm Sch}[h^{(1,1)}_{\alpha \beta}]_{\mu \nu}  &= 8\pi \alpha \beta T_{\mu \nu}^{(1,1)} 
+ 8\pi \alpha \beta \tilde{T}_{\mu \nu}^{(1,1)} 
=: 8\pi \alpha \beta T_{\mu \nu}^{{\rm eff}(1,1)}
\\
\alpha^2 \beta {\cal L}^{\rm Sch}[h^{(2,1)}_{\alpha \beta}]_{\mu \nu}  &= 
8\pi \alpha^2 \beta T_{\mu \nu}^{(2,1)}
+
8\pi \alpha^2 \beta  \tilde{T}_{\mu \nu}^{(2,1)}
=: 8\pi \alpha^2 \beta  T_{\mu \nu}^{{\rm eff}(2,1)}
\end{align}
If we regard $\alpha^n \beta~(n=0,1,2)$ as small parameters, we can apply the formalism in Sec.~\ref{sec:schperturbation} to
these equations at each order.

\subsection{The Eddington-Finkelstein like coordinates}

We discuss the backreaction of the Blandford-Znajek process using the formalism developed in Sec.~\ref{sec:schperturbation} which is written in the Eddington-Finkelstein coordinates.
Then, it is convenient to introduce the Eddington-Finkelstein like coordinates $(V,r,\theta,\Phi)$ by $dV = dT + dr$.
In this coordinate system, the Kerr metric becomes
\begin{align}
g_{\mu \nu}^{\rm EF} dx^\mu dx^\nu &= 
- \left(1-\frac{2Mr}{\Sigma}\right)dV^2 
+ 2 dV dr + \Sigma d\theta^2
+
\frac{(r^2 + a^2)^2 - a^2 \Delta \sin^2\theta}{\Sigma} \sin^2\theta d\Phi^2
\notag\\&
-
2 a \sin^2\theta dr d\Phi
-\frac{4 a M r}{\Sigma} \sin^2\theta dVd\Phi.
\label{metricKerrEF}
\end{align}
Then, $h_{\mu \nu}^{(n,0)}~(n=0,1,2,\cdots)$ in Eq.~\eqref{eq:hn0} 
can be obtained by taking a Taylor series around $a = 0$ 
for the metric Eq.~\eqref{metricKerrEF}, {\it i.e.,}
\begin{align}
g_{\mu \nu}^{\rm EF} dx^\mu dx^\nu &= 
(g_{\mu \nu}^{\rm Sch} + \alpha h_{\mu \nu}^{(1,0)} + \alpha^2 h_{\mu \nu}^{(2,0)} + {\cal O}(\alpha^3) )dx^\mu dx^\nu,
\end{align}
with
\begin{align}
g_{\mu \nu}^{\rm Sch} dx^\mu dx^\nu &=
-f dV^2 + 2 dV dr + r^2(d\theta^2 + \sin^2\theta d\Phi^2),
\\
\alpha h_{\mu \nu}^{(1,0)} dx^\mu dx^\nu &=
- 2 \alpha M \sin^2\theta d\Phi\left(
\frac{2M}{r}dV + dr
\right),
\label{hmunu10}
\\
\alpha^2 h_{\mu \nu}^{(2,0)} dx^\mu dx^\nu &=
\alpha^2 M^2 \left[
-  \frac{2 M }{r^3}\cos^2\theta dV^2
+ \cos^2\theta d\theta^2
+
 \frac{(r + M - M \cos(2\theta)) \sin^2\theta }{r}d\Phi^2
\right].
\end{align}

In the Eddington-Finkelstein like coordinates $(V,r,\theta,\Phi)$,
we obtain the following equations for the energy-momentum tensors of the Blandford-Znajek process
discussed in Sec.~\ref{sec:bz}:
\begin{align}
\beta T_{\mu \nu}^{(0,1)}dx^\mu dx^\nu &=  \frac{M^2 \beta }{2r^4} \left[
\left(1-\frac{2M}{r}\right)dV^2
-2 dVdr
+
r^2(d\theta^2 + \sin^2\theta d\Phi^2)
\right],
\label{tmunuBZ0}
\\
\alpha \beta T_{\mu \nu}^{(1,1)}dx^\mu dx^\nu &= 
- 2 \alpha \beta M \sin^2\theta d\Phi\left[
\frac{r^3 - 8M^3}{8r^5} dV
-
\frac{2M^2 + r(r+2M)}{4r^4} dr
\right],
\label{tmunuBZ1}
\\
\alpha^2 \beta  T_{\mu \nu}^{(2,1)}dx^\mu dx^\nu &= 
\alpha^2 \beta  T_{\mu \nu}^{(2,1)}|_{\ell = 0}dx^\mu dx^\nu 
+
\alpha^2 \beta  T_{\mu \nu}^{(2,1)}|_{\ell = 2}dx^\mu dx^\nu,
\end{align}
with
\begin{align}
& \alpha^2 \beta  T_{\mu \nu}^{(2,1)}|_{\ell = 0}dx^\mu dx^\nu  
\notag\\&= 
 \frac{ \alpha^2 \beta}{96 r^7}\bigg[
(96M^5 -64M^4r + 16 M^3 r^2 + r^5)dV^2
\notag\\& 
-4 r (4M^2 + r^2) (-4M^2 + 2Mr + r^2) dV dr
\notag\\& 
+4r^3(r+2M)^2 dr^2
-32  M^4(M+r) r^2(d\theta^2 +\sin^2\theta d\Phi^2)
\bigg],
\label{tmunuBZ2ell0}
\\
& \alpha^2 \beta  T_{\mu \nu}^{(2,1)}|_{\ell = 2}dx^\mu dx^\nu  
\notag\\&= 
 \frac{ \alpha^2 \beta}{48 r^7}\sqrt{\frac{\pi}{5}} Y_{2,0} \bigg[
\Big(192M^5 - r (32M^4 + 16M^3 r + r^4) + 192 M^2(-2M + r) r^2 {\cal F}\Big)dV^2
\notag\\&+
4  r (32M^4 + 8 M^3r + 2 Mr^3 + r^4 - 96M^2 r^2 {\cal F}) dVdr
- 4 r^3 (r+2M)^2dr^2
\notag\\&
+16   (2M^3 - rM^2 + 12 r^3 {\cal F})r^2(d\theta^2 +\sin^2\theta d\Phi^2)
\bigg]
- \sqrt{\frac{\pi}{5}} \frac{4  \alpha^2 \beta M^2 \partial_r{\cal F}}{3r^2}  (\partial_\theta Y_{2,0}) dr d\theta 
\notag\\&
+ \sqrt{\frac{\pi}{5}} \frac{ \alpha^2 \beta M^3(r^2 + rM + 2M^2)}{3r^5}
\sum_{i,j = \theta,\Phi}\left(\hat{\nabla}_i \hat{\nabla}_j Y_{2,0} - \frac{1}{2}\gamma_{ij}\hat{\triangle}Y_{2,0}\right)dx^i dx^j,
\label{tmunuBZ2ell2}
\end{align}
where $Y_{2,0} =4^{-1}\sqrt{5/\pi} (-1 + 3 \cos^2\theta)$,
$\gamma_{ij}$ is the metric of the unit sphere, {\it i.e.}, $\sum_{i,j=\theta,\Phi}\gamma_{ij}dx^idx^j = d\theta^2 + \sin^2\theta d\Phi^2$,
the operators
$\hat{\nabla}_i$ and $\hat{\triangle}$ denote the covariant derivative and the Laplacian on $\gamma_{ij}$, respectively.

\subsection{${\cal O}(\beta)$ corrections}
\label{sec:Obeta}
We can read ${\cal A}^{(0,1)}, T_{Vr}^{(0,1)}$ and $T_{rr}^{(0,1)}$ from the ${\cal O}(\beta^2)$ energy-momentum tensor in Eq.~\eqref{tmunuBZ0}
as
\begin{align}
{\cal A}^{(0,1)} = 0, \quad \beta T_{Vr}^{(0,1)} = - \frac{\beta M^2}{2r^4}, \quad T_{rr}^{(0,1)} = 0.
\end{align}
{}From Eqs.~\eqref{h0even},\eqref{h1even},\eqref{deltaMeven} and \eqref{lambdaeven}, we obtain the perturbed metric as
\begin{align}
\beta h^{(0,1)}_{\mu \nu} dx^\mu dx^\nu 
&= \beta \left[\frac{2\delta m^{(0,1)}}{r} + \frac{ 2\pi M(r-2M)}{r^2}\right]dV^2,
\end{align}
where we set the residual gauge mode as $\chi^{(0,1)}(V) = 0$.
For the later convenience, we choose $\delta m^{(0,1)} = -\pi M$, then the total metric at this order is
\begin{align}
(g_{\mu \nu}^{\rm Sch} + \beta h^{(0,1)}_{\mu \nu}) dx^\mu dx^\nu
&=
-\left(1 - \frac{2 M}{r} + \frac{4\pi M^2\beta}{r^2}\right) dV^2
+2 dVdr+ r^2(d\theta^2 + \sin^2\theta d\Phi^2).
\label{eq:magRNmetric}
\end{align}
This is the Reissner-Nordstr\"om metric with a magnetic charge parameter $Q = 2\sqrt{\pi \beta}M=2 \sqrt{\pi}C$.
We can see that the mass of the spacetime is $M$ and the location of the event horizon is $r = r_H$ with
\begin{align}
r_H = M + \sqrt{M^2 - 4\pi M^2\beta^2} = 2M - 2\pi M \beta  + {\cal O}(\beta^2).
\end{align}

One may think that it is strange for the spacetime to be the 
magnetic Reissner-Nordstr\"om metric because the Blandford-Znajek solution
is globally different from the magnetic monopole, but it describes the split monopole~\cite{Blandford:1977ds}.
The answer to this question is because the Birkhoff's theorem for the specially symmetric spacetime holds {\it locally},
and the energy-momentum tensor of the Blandford-Znajek solution at ${\cal O}(\beta)$ is locally same as that 
for the global magnetic monopole.

\subsection{${\cal O}(\alpha \beta)$ corrections: time dependence of the angular momentum}
After some calculations, we obtain 
\begin{align}
 8\pi \alpha \beta \tilde{T}_{\mu \nu}^{(1,1)} dx^\mu dx^\nu &= 
16 \pi \alpha \beta \sin^2\theta d\Phi\left(
\frac{ M^4 }{r^5} dV
+
 \frac{ M^3}{2r^4} dr
\right).
\end{align}
Thus, the effective energy-momentum tensor becomes
\begin{align}
 8\pi \alpha \beta T_{\mu \nu}^{{\rm eff}(1,1)} dx^\mu dx^\nu 
&=
 8\pi \alpha \beta T_{\mu \nu}^{(1,1)} dx^\mu dx^\nu
+
 8\pi \alpha \beta \tilde{T}_{\mu \nu}^{(1,1)} dx^\mu dx^\nu
\notag\\&=
-16 \pi \alpha \beta \sin^2\theta d\Phi\left[
 \frac{ M(r^3-16M^3)}{8r^5}dV
-
\frac{M(4M^2 + 2 M r + r^2)}{4r^4} dr
\right].
\end{align}
We can read 
\begin{align}
 t_{V\Phi}^{{\rm eff}(1,1)} &=  \frac{ M(r^3-16M^3)}{8r^5}, 
\\
t_{r\Phi}^{{\rm eff}(1,1)} &= -
\frac{M(4M^2 + 2 M r + r^2)}{4r^4},
\end{align}
then, ${\cal B}^{{\rm eff}(1,1)}$ in Eq.~\eqref{eq:defb} and $h_0^{{\rm IH}(1,1)}$ in Eq.~\eqref{eq:ihsol} become
\begin{align}
{\cal B}^{{\rm eff}(1,1)} &= - \frac{\pi}{3}, 
\\
 h_0^{{\rm IH}(1,1)} &= \frac{\pi r^2}{36M} 
   \left[-13 + \frac{8 M^2 \left(-18 M^2 + 4 M r + 9 r^2 - 12 M r \ln(2M/r)\right)}{r^4}\right].
\end{align}
We note that ${\cal B}^{{\rm eff}(1,1)}$ is constant because
$T_{\mu \nu}^{{\rm eff}(1,1)} $ does not have a time dependence (see Eq.~\eqref{eq:drb}).
From Eq.~(\ref{eq:h-l1}), 
the perturbed metric becomes
\begin{align}
\alpha \beta h_{\mu \nu}^{(1,1)} dx^\mu dx^\nu
&=
-
\frac{4 M \alpha \beta \sin^2\theta}{r} d\Phi dV 
\left[
\delta a^{(1,1)}
+ 
{\cal B}^{{\rm eff}(1,1)}(V - V_0) + \frac{r}{2M}\Big(
h_0^{{\rm IH}(1,1)} + r^2 C_2^{(1,1)}(V)
\Big)
\right].
\label{eq:metric11}
\end{align}
At this order, the Komar angular momentum at the radius $r$ is
\begin{align}
J_{\rm Komar} = \alpha M^2 + 
\alpha \beta M \left[
\delta a^{(1,1)} + {\cal B}^{{\rm eff}(1,1)}(V - V_0)
+ \frac{r}{6M}\left(2 h_0^{{\rm IH}(1,1)} - r \partial_r h_0^{{\rm IH}(1,1)} \right)
\right].
\label{eq:jkomar11anyr}
\end{align}
We find that the time dependence of $J_{\rm Komar}$ coincides with the prediction from 
the angular momentum extraction rate of the Blandford-Znajek process in Eq.~\eqref{eq:amexrateBZ}
\begin{align}
\partial_V J_{\rm Komar} &= \alpha \beta M {\cal B}^{{\rm eff}(1,1)} = - \frac{\alpha \beta \pi M }{3}.
\end{align}
We set $\delta a^{(1,1)} = 0$, then $J_{\rm Komar} = \alpha M^2$ at $V = V_0$ and $r = r_0$.
We also choose the gauge mode as $C_2^{(1,1)}(V) = 13 \pi/(36 M)$ so that the divergent behavior $h_0^{{\rm IH}(1,1)}$ at $r \to \infty$
is canceled in Eq.~\eqref{eq:metric11}.

\subsection{${\cal O}(\alpha^2 \beta)$ corrections: time dependence of the mass}
In a similar way, we obtain $\tilde{T}_{\mu \nu}^{(2,1)}$ as
\begin{align}
8\pi \alpha^2 \beta  \tilde{T}_{\mu \nu}^{(2,1)} 
&=
8\pi \alpha^2 \beta \tilde{T}_{\mu \nu}^{(2,1)}|_{\ell = 0}
+
8\pi \alpha^2 \beta  \tilde{T}_{\mu \nu}^{(2,1)}|_{\ell = 2},
\end{align}
with
\begin{align}
& 8\pi \alpha^2 \beta  \tilde{T}_{\mu \nu}^{(2,1)}|_{\ell = 0}dx^\mu dx^\nu  
\notag\\&= 
\frac{\alpha^2\beta}{27 r^8}\bigg[
\Big(
-\pi (288 M^6 - 936 M^5 r + 288 M^4 r^2 + 144 M^3 r^3 + 26 M r^5 - 
    13 r^6) 
\notag\\&
+ 
 {\cal B}^{{\rm eff}(1,1)} (72 M^2 r^3 (-3 M + r) - 36 M^2 (9 M - 4 r) (2 M - r) r(V-V_0)) 
\notag\\&
+ 
 18 M (2 M - 
    r) r^2 (2 (-3 M + r) h_0^{{\rm IH}(1,1)}  + (3 M - 2 r) r \partial_r h_0^{{\rm IH}(1,1)} )
\Big) dV^2
\notag\\& 
+2 r \Big(
\pi (-144 M^5 + 432 M^4 r + 72 M^3 r^2 - 36 M^2 r^3 - 13 r^5) 
\notag\\&
+ 
 {\cal B}^{{\rm eff}(1,1)} (-54 M^2 r^3 + 36 M^2 r (-9 M + 4 r)(V-V_0)) 
\notag\\&
+ 
 18 M r^2 (2 (-3 M + r) h_0^{{\rm IH}(1,1)}  + (3 M - 2 r) r \partial_r h_0^{{\rm IH}(1,1)} )
\Big) dVdr
\notag\\& 
+ 36 r^2 \Big(
-6 {\cal B}^{{\rm eff}(1,1)} M^2 r(V-V_0) + 
 M (2 M \pi (-2 M^2 + 2 M r + r^2) - 2 r^2 h_0^{{\rm IH}(1,1)}  + 
    r^3 \partial_r h_0^{{\rm IH}(1,1)} )
\Big) dr^2
\notag\\&
+18 M r^3 \Big(
4 M^2 \pi (-3 M^2 + M r + 2 r^2) 
+ 
 {\cal B}^{{\rm eff}(1,1)} (-6 M r^3 + 6 M r (-3 M + r)(V-V_0)) 
\notag\\&
+ (3 M - r) r^2 (-2 h_0^{{\rm IH}(1,1)}  + 
    r \partial_r h_0^{{\rm IH}(1,1)} )
\Big)(d\theta^2 +\sin^2\theta d\Phi^2)
\bigg],
\end{align}
and
\begin{align}
& 8\pi a^2 C^2 \tilde{T}_{\mu \nu}^{(2,1)}|_{\ell = 2}dx^\mu dx^\nu  
\notag\\&  = 
\frac{4\alpha^2 \beta}{27 r^8}\sqrt{\frac{\pi}{5}} Y_{2,0} \bigg[
\Big(
\pi (144 M^6 - 792 M^5 r + 144 M^4 r^2 + 72 M^3 r^3 + 13 M r^5 - 
    26 r^6) 
\notag\\&
+ 
 {\cal B}^{{\rm eff}(1,1)} (36 M^2 (3 M - r) r^3 + 18 M^2 r (18 M^2 - 17 M r - 2 r^2)(V-V_0)) 
\notag\\&
+ 
 9 M r^2 (2 (6 M^2 - 5 M r - 2 r^2) h_0^{{\rm IH}(1,1)}  - 
    r (-2 M + r) (-3 M + 2 r) \partial_r h_0^{{\rm IH}(1,1)} )
\Big)dV^2
\notag\\&+
\frac{r}{2} \Big(
\pi (288 M^5 - 864 M^4 r - 144 M^3 r^2 + 72 M^2 r^3 + 65 r^5) 
\notag\\&
+ 
 {\cal B}^{{\rm eff}(1,1)} (108 M^2 r^3 + 72 M^2 (9 M - r) r(V-V_0)) 
\notag\\&
+ 
 36 M r^2 ((6 M + r) h_0^{{\rm IH}(1,1)}  + r (-3 M + 2 r) \partial_r h_0^{{\rm IH}(1,1)} )
\Big) dVdr
\notag\\&
+18 Mr^2 \Big(
2 M \pi (2 M^2 - 2 M r - r^2) + 6 {\cal B}^{{\rm eff}(1,1)} M r(V-V_0) + 2 r^2 h_0^{{\rm IH}(1,1)}  - 
 r^3 \partial_r h_0^{{\rm IH}(1,1)} 
\Big)dr^2
\notag\\&
+\frac{3r^3}{4} \Big(
\pi (144 M^5 - 264 M^4 r - 96 M^3 r^2 + 13 r^5) + 
 {\cal B}^{{\rm eff}(1,1)} (72 M^2 r^3 + 216 M^3 r(V-V_0)) 
\notag\\&
+ 
 12 M r^2 ((6 M + r) h_0^{{\rm IH}(1,1)}  + r (-3 M + r) \partial_r h_0^{{\rm IH}(1,1)} )
\Big)(d\theta^2 +\sin^2\theta d\Phi^2)
\bigg]
\notag\\&
+2\sqrt{\frac{\pi}{5}} \frac{\alpha^2 \beta M}{27 r^6} 
\Big(
720 M^4 \pi + 13 \pi r^4 + {\cal B}^{{\rm eff}(1,1)} (36 M r^3 + 36 M r (8 M + 3 r)(V-V_0)) 
\notag\\&
+ 
 36 r^2 (3 M + r) h_0^{{\rm IH}(1,1)}  - 18 r^3 (2 M + r) \partial_r h_0^{{\rm IH}(1,1)} 
\Big) (\partial_\theta Y_{2,0}) dV d\theta 
\notag\\&
+2 \sqrt{\frac{\pi}{5}}\frac{\alpha^2\beta}{27 r^4}
\Big(-180{\cal B}^{{\rm eff}(1,1)} M^2(V-V_0) 
\notag\\&
+ 
 r (-72 M h_0^{{\rm IH}(1,1)}  + r (-13 \pi r + 18 M \partial_r h_0^{{\rm IH}(1,1)} ))
\Big) (\partial_\theta Y_{2,0}) dr d\theta 
\notag\\&
+ \sqrt{\frac{\pi}{5}} 
\frac{\alpha^2 \beta }{27 r^5}
\Big(
\pi (-720 M^5 - 72 M^4 r + 13 r^5) + 72 {\cal B}^{{\rm eff}(1,1)} M^2 r (-9 M + r)(V-V_0) 
\notag\\&
+ 
 36 M r^2 ((-6 M + r) h_0^{{\rm IH}(1,1)}  + 3 M r \partial_r h_0^{{\rm IH}(1,1)} )
\Big)
\sum_{i,j = \theta,\Phi}\left(\hat{\nabla}_i \hat{\nabla}_j Y_{2,0} - \frac{1}{2}\gamma_{ij}\hat{\triangle}Y_{2,0}\right)dx^i dx^j.
\end{align}
The effective energy-momentum tensor $T_{\mu \nu}^{{\rm eff}(2,1)}$ is given by
\begin{align}
8\pi \alpha^2 \beta  T_{\mu \nu}^{{\rm eff}(2,1)}
&=
8\pi  \alpha^2 \beta  T_{\mu \nu}^{(2,1)}
+
8\pi  \alpha^2 \beta  \tilde{T}_{\mu \nu}^{(2,1)}.
\end{align}
{}From the expression of $T_{\mu \nu}^{{\rm eff}(2,1)}|_{\ell = 0}$, we can read ${\cal A}^{{\rm eff}(2,1)}, T_{Vr}^{{\rm eff}(2,1)}$ and $T_{rr}^{{\rm eff}(2,1)}$ as
\begin{align}
{\cal A}^{{\rm eff}(2,1)} &= \frac{1}{24 r^3}\Big[8M^2{\cal B}^{{\rm eff}(1,1)} (r-6M) 
- \pi r(8M^2 + r^2)
\Big],
\label{eq:aeff21}
\\
\alpha^2\beta T_{Vr}^{{\rm eff}(2,1)} &= \frac{\alpha^2 \beta}{432 \pi r^7}\Big[
-\pi (288 M^5 - 1008 M^4 r - 72 M^3 r^2 + 72 M^2 r^3 + 18 M r^4 + 
    35 r^5) 
\notag\\&
+ {\cal B}^{{\rm eff}(1,1)} (-108 M^2 r^3 + 72 M^2 r (-9 M + 4 r)(V-V_0)) 
\notag\\&
+ 
 36 M r^2 (2 (-3 M + r) h_0^{{\rm IH}(1,1)} + (3 M - 2 r) r \partial_r h_0^{{\rm IH}(1,1)} )
\Big]
\\
\alpha^2\beta T_{rr}^{{\rm eff}(2,1)} &= \frac{\alpha^2 \beta}{24 \pi r^6} \Big[
\pi (-16 M^4 + 16 M^3 r + 12 M^2 r^2 + 4 M r^3 + r^4) 
\notag\\&
- 
 24 {\cal B}^{{\rm eff}(1,1)} M^2 r(V-V_0) + 4 M r^2 (-2 h_0^{{\rm IH}(1,1)} + r \partial_r h_0^{{\rm IH}(1,1)})
\Big].
\end{align}
We can see that ${\cal A}^{{\rm eff}(2,1)}$ is not a constant but does not depend on $V$.
The value of ${\cal A}^{{\rm eff}(2,1)}$ at $r = r_0$ is
\begin{align}
{\cal A}^{{\rm eff}(2,1)}|_{r = r_0} =  - \frac{5\pi}{72}.
\label{eq:aeffeqhorizon}
\end{align}
{}From Eqs.~\eqref{h0even},\eqref{h1even},\eqref{deltaMeven} and \eqref{lambdaeven}, we obtain the perturbed metric as
\begin{align}
\alpha^2 \beta  h_{\mu \nu}^{(2,1)}|_{\ell=0} dx^\mu dx^\nu
&=
\alpha^2 \beta  \left[\left(\frac{2 \delta M^{(2,1)}}{r} + 2 f \lambda^{(2,1)} \right)dV^2 - 2 \lambda^{(2,1)} dVdr 
\right]
\end{align}
with
\begin{align}
\delta M^{(2,1)} &= \delta m^{(2,1)} + {\cal A}^{{\rm eff}(2,1)}(V-V_0) 
- 4\pi \int_{r_0}^r \bar{r}^2 T_{Vr}^{{\rm eff}(2,1)}(V_0,\bar{r}) d\bar{r},
\label{eq:deltaM21}
\\
\lambda^{(2,1)} &= - 4\pi \int_{r_0}^r \bar{r} T_{rr}^{{\rm eff}(2,1)}(V,\bar{r}) d\bar{r} + \chi^{(2,1)}(V),
\end{align}
where $\delta m^{(2,1)}$ is a constant and the function $\chi^{(2,1)}(V)$ corresponds to the residual gauge mode.
We can see that ``the mass term'' $\delta M^{(2,1)}$ depends on time.
However, because the spacetime is not spherically symmetric at this order, 
the appropriate definition of the mass is not clear. We discuss this topic in the next section.

We should note that $\ell = 2$ even-parity metric perturbations also exist at ${\cal O}(\alpha^2 \beta)$
\begin{align}
\alpha^2 \beta  h_{\mu \nu}^{(+)(2,1)}|_{\ell=2} dx^\mu dx^\nu
&=
4 \alpha^2 \beta \sqrt{\frac{\pi}{5}}Y_{2,0} 
\Big[
H_{0, \ell =2}^{(2,1)}dV^2
+
2 H_{1, \ell =2}^{(2,1)}dVdr
+
H_{2, \ell =2}^{(2,1)} dr^2
\notag\\
&
+2 K_{\ell =2}^{(2,1)} r^2(d\theta^2 + \sin^2\theta d\Phi^2)
\Big].
\end{align}
The perturbed metric can be obtained by solving the Zerilli equation~\cite{Zerilli:1970se, Zerilli:1971wd, Martel:2005ir}.
As shown in the next section, $\ell = 2$ metric perturbations do not affect the area of the apparent horizon,
and thus these modes are not relevant for the discussion of the black hole mechanics.

\section{Black Hole Mechanics}
\label{sec:bhmechanics}

In this section, we discuss the 
relation among the area, mass and angular momentum of the black hole.
In the standard derivation of the black hole mechanics~\cite{Bardeen:1973gs, Wald:1984rg}, 
assuming the time-translational and rotational Killing vectors in a vacuum spacetime before and 
after the dynamical process, the differences in the Bondi-Sachs energy and angular momentum and therefore the energy and angular momentum of the whole system are  discussed. In the present case, however, we would like to determine the energy and angular momentum extraction in the presence of the force-free electromagnetic fields without a time-translational Killing vector.
We do not assume the stationary stages before and after the energy extraction. Moreover, to isolate the energy and angular momentum of the black hole from the ambient electromagnetic fields, we need to discuss them in terms of quasi-local quantities.
In the present situation, we show that the apparent horizon is a good candidate for the black hole horizon for this purpose and that the first law of black hole mechanics holds if we take the appropriate time-dependent mass parameter of the apparent horizon.

\subsection{Apparent Horizon}
In this subsection, we discuss the apparent horizon for the metric $g_{\mu \nu} = g_{\mu \nu}^{\rm Kerr} +g_{\mu \nu}^{\rm BZ}$. Because $V = {\rm const.}$ surface of this spacetime is timelike at ${\cal O}(\alpha^2 \beta)$,
we work in the Kerr Schild coordinates $(T,r,\theta, \Phi)$.
We set the relation between $T$ and $V$ as $V = T + r-2M$.
The unit normal to $T = {\rm const.}$ surface is given by
\begin{align}
n_\mu dx^\mu = F_n dT,
\end{align}
where the function $F_n$ is chosen so that $g^{\mu \nu}n_\mu n_\nu = -1$ and $n^\mu$ is future directed.
The induced metric on $T = {\rm const.}$ surface is given by
\begin{align}
\gamma_{\mu \nu} = g_{\mu \nu} + n_\mu n_\nu,
\end{align}
and the projection operator $\gamma_{\mu}{}^{\nu}$ becomes
\begin{align}
\gamma_{\mu}{}^{\nu} = \gamma_{\mu \alpha}g^{\alpha \nu}.
\end{align}
Because of the facts
\\
$\bullet$ $Y_{0,0}$ perturbations come from ${\cal O}(\beta)$ and ${\cal O}(\alpha^2\beta)$,
\\
$\bullet$ $Y_{1,0}$ perturbations come from ${\cal O}(\alpha)$ and ${\cal O}(\alpha \beta)$,
\\
$\bullet$ $Y_{2,0}$ perturbations come from ${\cal O}(\alpha^2)$ and ${\cal O}(\alpha^2\beta)$,
\\
in the metric $g_{\mu \nu} = g_{\mu \nu}^{\rm Kerr} +g_{\mu \nu}^{\rm BZ}$,
we can assume that the location of the apparent horizon at each hypersurface $T = {\rm const.}$ is
\begin{align}
r &=  {\cal R}(\theta;T)
\notag\\&= ({\cal R}^{(0,0)} + \beta {\cal R}^{(0,1)} + \alpha^2 {\cal R}^{(2,0)}_{\ell = 0}+ \alpha^2 \beta {\cal R}^{(2,1)}_{\ell = 0}) 
\notag\\& + \alpha ({\cal R}^{(1,0)} + \beta {\cal R}^{(1,1)}) 2\sqrt{\frac{\pi}{3}} Y_{1,0} 
+ \alpha^2 ({\cal R}^{(2,0)}_{\ell = 2} + \beta {\cal R}^{(2,1)}_{\ell = 2}) 4\sqrt{\frac{\pi}{5}}  Y_{2,0},
\label{eq:ahorizon}
\end{align}
where the coefficients 
only depend on $T$~\cite{Nakamura:1987zz}.
{}From the results for the Kerr metric
and ${\cal O}(\beta)$ perturbations, 
where the location of the apparent horizon 
coincides with that of the event horizon at this order, discussed in Sec.~\ref{sec:Obeta},
 we obtain
\begin{align}
{\cal R}^{(0,0)}  &= 2 M,
\\
\alpha {\cal R}^{(1,0)}&= 0,
\\
\alpha^2 {\cal R}^{(2,0)}_{\ell = 0}&= - \frac{\alpha^2 M}{2},
\\
\alpha^2 {\cal R}^{(2,0)}_{\ell = 2} &= 0,
\\
\beta {\cal R}^{(0,1)} &= - 2\pi \beta M.
\end{align}
Thus, we need to fix ${\cal R}^{(2,1)}_{\ell = 0}, {\cal R}^{(1,1)}, {\cal R}^{(2,1)}_{\ell = 2}$.
The unit normal to $r =  {\cal R}(\theta;T)$ at each $T = {\rm const}$ surface is
\begin{align}
s_\mu = F_s \gamma_\mu{}^\nu \bar{s}_\nu,
\end{align}
with $\bar{s}_\mu dx^\mu = dr - (\partial_{\theta}  {\cal R)} d\theta$,
where the function $F_s$ is chosen so that $g^{\mu \nu}s_\mu s_\nu= 1$ and $s^\mu$ is an outward vector.
The induced metric on the $T = {\rm const.}$ and $r =  {\cal R}(\theta;T)$ surface is
\begin{align}
q_{\mu \nu} = \gamma_{\mu \nu} - s_\mu s_\nu = g_{\mu \nu} + n_\mu n_\nu - s_\mu s_\nu.
\end{align}
The location of the apparent horizon is determined by the equation
\begin{align}
\theta_+ = q^{\mu \nu}\nabla_\mu (n_\nu + s_\nu) = 0.
\label{eq:thetaplus}
\end{align}
After some calculations, we obtain
\begin{align}
{\cal R}^{(1,1)} &= 0,
\\
{\cal R}^{(2,1)}_{\ell = 0} &=
\left(2{\cal A}^{{\rm eff}(2,1)}|_{r=r_0} 
- \frac{2}{3}{\cal B}^{{\rm eff}(1,1)}
\right) (T-T_0)
- \frac{151 M\pi}{54}
+ 2 \delta m^{(2,1)},
\\
{\cal R}^{(2,1)}_{\ell = 2} &= 
\frac{{\cal B}^{{\rm eff}(1,1)}}{21}(T-T_0)
+
\frac{5 \pi M}{54} 
+ 
\frac{2 M}{7} H_{0, \ell =2}^{(2,1) }|_{r=r_0} 
- 
\frac{8 M^2}{7} \partial_T K^{(2,1)}_{\ell =2}|_{r=r_0} ,
\end{align}
as solutions of Eq.~\eqref{eq:thetaplus}, where $T_0 = V_0$.
Using our results in the previous section, we have a relation
\begin{align}
2{\cal A}^{{\rm eff}(2,1)}|_{r=r_0} 
- \frac{2}{3}{\cal B}^{{\rm eff}(1,1)} = \frac{\pi}{12}.
\end{align}
The area of the apparent horizon is given by
\begin{align}
A_{\rm AH} &= 
16 \pi M^2 - 32 \pi M^2\beta - 4 \pi M^2 \alpha^2 + \frac{4\pi M \alpha^2 \beta}{3}\Big[
7 \pi M + 12 {\cal R}^{(2,1)}_{\ell = 0}
\Big] + {\cal O}(\alpha^3) + {\cal O}(\beta^2).
\end{align}
We should note that $\ell =2$ terms in Eq.~\eqref{eq:ahorizon} do not affect 
the area because of the orthogonality of the spherical harmonics.
Thus, the time dependence of the apparent horizon area is
\begin{align}
\partial_T A_{\rm AH} &= 16 \pi M \alpha^2 \beta \partial_T {\cal R}^{(2,1)}_{\ell = 0}
\notag\\&=
\frac{4 \pi^2 M \alpha^2 \beta}{3}.
\label{eq:dtaah}
\end{align}

\subsection{Angular Momentum}

The Komar angular momentum at the apparent horizon is
\begin{align}
J_{\rm Komar}|_{\rm AH} = \alpha M^2 + 
\alpha \beta M  {\cal B}^{{\rm eff}(1,1)}(T - T_0).
\label{eq:jkomar11}
\end{align}
The time dependence of $J_{\rm Komar}|_{\rm AH}$ is
\begin{align}
\partial_T J_{\rm Komar}|_{\rm AH} &= \alpha \beta M {\cal B}^{{\rm eff}(1,1)}
\notag\\&= 
- \frac{\alpha \beta \pi M }{3}
\notag\\&= - \dot{J}_{\rm BZ},
\label{eq:dtjkomar}
\end{align}
where $\dot{J}_{\rm BZ}$ is given by Eq.~\eqref{eq:amexrateBZ}.
Thus, this 
reproduces 
the angular momentum extraction rate of the Blandford-Znajek process in Eq.~\eqref{eq:amexrateBZ}.
This explicitly shows that the total angular momentum conservation law holds, {\it i.e.,}
the decreasing rate of the angular momentum of the black hole
is balanced with the angular momentum extraction rate of the Blandford-Znajek process.

\subsection{Implication from the Black Hole Mechanics}
If we assume the relation of the first law of black hole mechanics~\cite{Bardeen:1973gs}
\begin{align}
d M = \frac{\kappa}{8\pi}d A + \Omega_H dJ,
\end{align}
we can obtain implication of the time dependence of the black hole mass.
Setting $dA$ and $dJ$ as $\partial_T A_{\rm AH}$ and $\partial_T J_{\rm Komar}|_{\rm AH}$ 
in Eqs.~\eqref{eq:dtaah} and \eqref{eq:dtjkomar},
the time dependence of the mass is suggested by
\begin{align}
\partial_T M = 
\frac{\kappa}{8\pi}\frac{4 \pi^2 M \alpha^2 \beta}{3} + \Omega_H \left(- \frac{\alpha \beta \pi M }{3}\right).
\end{align}
If we assume
\begin{align}
\kappa &= \frac{1}{4 M} + {\cal O}(\alpha) + {\cal O}(\beta),
\\
\Omega_H &= \frac{\alpha}{4 M} + {\cal O}(\alpha \beta) + {\cal O}(\alpha^2) + {\cal O}(\beta),
\end{align}
we obtain
\begin{align}
\partial_T M &= - \frac{\alpha^2 \beta \pi}{24},
\notag\\&=-\dot{E}_{\rm BZ},
\label{eq:partialtm}
\end{align}
where $\dot{E}_{\rm BZ}$ is given by Eq.~\eqref{eq:energyexrateBZ}.
This reproduces  
the energy extraction rate of the Blandford-Znajek process in Eq.~\eqref{eq:amexrateBZ},
although the quantity $M$ in the first law is yet undefined as a quasi-local quantity of the apparent horizon.

\subsection{The Hawking Mass}
Because the spacetime at ${\cal O}(\alpha^2 \beta)$ is not stationary nor spherically symmetric, 
it is not obvious how to define the mass of the black hole.
In that case, a possible choice of the quasi local mass is the Hawking mass.
The Hawking mass at the apparent horizon is given by~\cite{Hawking:1968qt, Jaramillo:2010ay}
\begin{align}
M_{\rm Hawking}|_{\rm AH} &= \sqrt{\frac{A_{\rm AH}}{16 \pi}},
\notag\\&=
M - \pi \beta M - \frac{\alpha^2 M}{8}
+
\frac{\alpha^2 \beta}{6}(\pi M + 3 {\cal R}^{(2,1)}_{\ell = 0})
+{\cal O}(\alpha^3) + {\cal O}(\beta^2).
\end{align}
The time dependence of $M_{\rm Hawking}|_{\rm AH}$ is
\begin{align}
\partial_T M_{\rm Hawking}|_{\rm AH} 
&= \frac{\alpha^2 \beta}{2} \partial_T {\cal R}^{(2,1)}_{\ell = 0}
\notag\\&=
\frac{\pi \alpha^2 \beta}{24} > 0.
\end{align}
While the absolute value is the desired value, this is positive.
This is because the Hawking mass at the apparent horizon is the square root of the apparent horizon area which is increasing in time. Thus, we consider that the Hawking mass is not suitable for the description of energy extraction 
by the Blandford-Znajek process.
We note that even if we use the Hayward mass~\cite{Hayward:1993ph, Jaramillo:2010ay}, 
the mass is increasing in time.

\subsection{Comparison with the Kerr Metric with Time-Dependent Parameters}

As shown in Appendix~\ref{appendix:Kerr}, the Kerr metric with the small parameter shifts
of the mass and angular momentum
takes the form of Eq.~\eqref{eq:metricKerralphabeta}.
In this subsection, we show that the time dependence of 
$g_{\mu\nu} = g_{\mu \nu}^{\rm Kerr} + g_{\mu \nu}^{\rm BZ}$
can be understood in terms of the Kerr metric 
Eq.~\eqref{eq:metricKerralphabeta}
but with time-decreasing mass and angular momentum.

Let us consider the Kerr metric in the form of Eq.~\eqref{eq:metricKerralphabeta}, 
but we replace the constants $\delta M^{\rm (phys)}$ and $\delta J^{\rm (phys)}$ 
by $\delta \bar{M}(V)$ and $\delta \bar{J}(V)$ which are the functions of $V$.
We denote by
$g^{{\rm Kerr}+(\delta \bar{M},\delta \bar{J})}_{\mu \nu}$
this metric.
We would like to compare $g_{\mu\nu} = g_{\mu \nu}^{\rm Kerr} + g_{\mu \nu}^{\rm BZ}$
with $g^{{\rm Kerr}+(\delta \bar{M},\delta \bar{J})}_{\mu \nu}$.
We set $\delta \bar{M}$ and $\delta \bar{J}$ as
\begin{align}
\alpha^2 \beta \delta \bar{M}
&= 
- \frac{\alpha^2 \beta \pi}{24}(V-V_0) = -\dot{E}_{\rm BZ}(V-V_0), 
\\
 \alpha \beta  \delta \bar{J} &=
- \frac{\alpha \beta \pi M }{3}(V-V_0) = -\dot{J}_{\rm BZ}(V-V_0).
\end{align}
We also choose $\chi(V)$ in $g^{{\rm Kerr}+(\delta \bar{M},\delta \bar{J})}_{\mu \nu}$ as (see Eq.~\eqref{eq:appendixchi})
\begin{align}
\chi = \chi^{(2,1)} +  \frac{1}{6M}{\cal B}^{{\rm eff}(1,1)}(V-V_0).
\end{align}
Then, after some calculations, we obtain 
\begin{align}
g_{\mu \nu}^{\rm Kerr} + g_{\mu \nu}^{\rm BZ} 
&= 
g^{{\rm Kerr}+(\delta \bar{M},\delta \bar{J})}_{\mu \nu}
+
g_{\mu \nu}^{\rm other}
+
[\ell = 2 ~{\rm terms}]
+ {\cal O}(\alpha^3)
+{\cal O}(\beta^2),
\label{eq:kerrsquence}
\end{align}
where $g_{\mu \nu}^{\rm other}$ does not depend on time.
We note that $g_{\mu \nu}^{\rm other}$ contains the ${\cal O}(\beta)$ effect, {\it i.e.,} 
the perturbation corresponding to the magnetic Reissner-Nordstr\"om metric duscussed in Sec.~\ref{sec:Obeta}.
Eq.~\eqref{eq:kerrsquence} shows that the $\ell =0, 1$ time-dependent terms of $g_{\mu\nu} = g_{\mu \nu}^{\rm Kerr} + g_{\mu \nu}^{\rm BZ}$
can be expressed as the Kerr metric with time-dependent parameters, 
$g^{{\rm Kerr}+(\delta \bar{M},\delta \bar{J})}_{\mu \nu}$, 
whose time dependence is determined from the energy and angular momentum extraction rates of the Blandford-Znajek process.
If we regard $M + \delta \bar{M}$ as a black hole mass, its time dependence coincides with Eq.~\eqref{eq:partialtm}.
Therefore, this gives an appropriate time-dependent mass for the energy extraction in the present setting.

We have seen that the $\ell =0, 1$ time-dependent terms of our results
can be fit by the Kerr metric with time-dependent parameters in
the Eddington-Finkelstein like coordinates.
We should note that it is essential on which time coordinate we let the mass and angular momentum parameters depend.
For example, if we let them depend on time in
the Boyer-Lindquist coordinates, our results cannot be fit by the corresponding spacetime.
Finding an appropriate coordinate system is not such a trivial problem, and 
what we showed is that the Eddington-Finkelstein like coordinates is the appropriate choice.\footnote{
In~\cite{Bohmer:2017sdz}, as an extension of the Vaidya metric~\cite{Vaidya:1951zz},
the Kerr metric but the time-dependent mass and angular momentum parameters 
are discussed in a different coordinate system from our paper.
It is interesting to discuss the relation with our perturbative solution, but we leave this problem for future work.}

Finally, we comment on ${\cal A}^{{\rm eff}(2,1)}$ in Eq.~\eqref{eq:aeff21}.
While ${\cal A}^{{\rm eff}(2,1)}$ is related to the flux associated with $\nabla^\mu(T_{\mu \nu}^{\rm eff}(\partial_V)^\nu) = 0$ (see Sec.~\ref{subsec:conservationlaw}), at this stage, the physical meaning of ${\cal A}^{{\rm eff}(2,1)}$ is not clear.
Because $\delta M^{(2,1)}$ in Eq.~\eqref{eq:deltaM21} is written by ${\cal A}^{{\rm eff}(2,1)}$,
it is useful to consider the meaning of ``the mass term'' $\delta M^{(2,1)}$.
If we compare the situation with the Kerr black hole case,
$\delta M^{(2,1)}$ corresponds to $\delta M^{(2,1)}_{\rm Kerr}$ in Eq.~\eqref{eq:appendixdeltaM21}.
As shown in Eq.~\eqref{eq:relationmphys},
``the mass term'' $\delta M^{(2,1)}_{\rm Kerr}$ in the Kerr black hole case
does not directly denote the variation of the mass of the black hole,
and the variation of the physical mass 
is obtained by subtracting the effect of the spin from $\delta M^{(2,1)}_{\rm Kerr}$, 
the second term of the right-hand side of Eq.~\eqref{eq:relationmphys}.
This suggests that $\delta M^{(2,1)}$ also contains information of both the mass and angular momentum of black holes,
and this is the reason why ${\cal A}^{{\rm eff}(2,1)}|_{r=r_0}$ in Eq.~\eqref{eq:aeffeqhorizon} does not coincide with
$-\dot{E}_{\rm BZ}$.
In fact, $ {\cal A}^{\rm eff(2,1)}$ can be written in terms of $\dot{E}_{\rm BZ}$ and $\dot{J}_{\rm BZ}$ as
\begin{align}
\alpha^2 \beta {\cal A}^{\rm eff(2,1)}= -\dot{E}_{\rm BZ} - \frac{4 M r - 6 M^2}{3 r^3} \alpha \dot{J}_{\rm BZ}.
\end{align}

\section{Summary and discussion}
\label{sec:summary}

We developed the formalism of 
monopole and dipole linear gravitational perturbations around the Schwarzschild black holes in the Eddington-Finkelstein coordinates against the generic time-dependent accreting matters.
We derived the mass and angular momentum of black holes in terms of the energy-momentum tensors of accreting matters, at the linear order.
The time dependence of the  mass and angular momentum are determined by the 
the accretion rates of the energy and angular momentum.
In particular, after the accreting matters completely fall into the black hole at some finite time,
$\ell = 0$ and $\ell = 1$ perturbations represent the slowly rotating Kerr black holes,
and the final mass and angular momentum are expressed by the total time integral of the accretion rates at $r = 2M$.
We also showed that our formalism can reproduce the exact Vaidya solution~\cite{Vaidya:1951zz}.

Applying our formalism to the Blandford-Znajek process~\cite{Blandford:1977ds},
we studied the metric backreaction.
While we need to 
study the non-linear gravitational perturbations
to discuss the backreaction of the Blandford-Znajek process, 
our formalism can be applied to this problem
because the forms of equations at each order are same as those of linear order 
with the source terms which contain the non-linear effects.
We calculated the time-dependent Komar angular momentum and area of the apparent horizon. The decreasing rate of the former coincides with the angular momentum loss rate estimated 
in terms of the stress-energy tensor of the force-free electromagnetic fields at infinity.

According to the test-field calculation of the energy and angular momentum extraction rates 
of the Blandford-Znajek process~\cite{Blandford:1977ds},
there is no doubt that energy and angular momentum are transfered to asymptotic regions.
However, it is not clear how to describe the local metric behavior of the backreaction.
In this paper, we showed that the time dependence of $\ell =0, 1$ modes are expressed by
the Kerr metric but with time-decreasing mass and angular momentum parameters, 
which depend only on the ingoing null coordinate $V$.
This suggests that the corresponding outgoing fluxes come directly from 
the vicinity of the event horizon.
If we regard the corresponding mass parameter as the black hole mass, we saw that its decreasing rate coincides with the energy extraction rate of the Blandford-Znajek process and that the first law of black hole mechanics holds for the apparent horizon 
in terms of this mass parameter but not the Hawking mass.

Finally, we comment on future works.
It is interesting to extend our analysis to 
the higher-order solutions of the Blandford-Znajek process~\cite{Tanabe:2008wm, Pan:2015iaa, Armas:2020mio}.
The applications to other situations, {\it e.g.,} the Penrose process or the superradiance phenomena,
is possible.
It is also interesting to consider the applications to modified gravity theories.
If we consider some modified gravity theories
and they admit solutions close to the Schwarzschild black holes,
we expect that the field equations for the monopole and dipole gravitational perturbations 
take the same form as Eq.~\eqref{perturbationeqsch}, then our formalism can be applied.

\begin{acknowledgments}
This work is partly supported by MEXT Grants-in-Aid for Scientific Research/JSPS KAKENHI Grants Nos. 20H04746 (M.K.), JP19K03876, JP19H01895, 20H05853 (T.H.), JP19H01891, 20H05852 (A.N.), and 18H01245 (K.T.).
\end{acknowledgments}

\appendix
\section{The gauge transformation of monopole and odd-parity dipole perturbations}
\label{appendix:gaugetr}
\subsection{$\ell = 0$ perturbations}
The general perturbed metric for the $\ell = 0$ linear perturbations in the Eddington-Finkelstein coordinates is given by
\begin{align}
 h_{\mu \nu}^{(+)} dx^\mu dx^\nu
&=
H_{0}(V,r)dV^2
+
2 H_{1}(V,r)dVdr
+
H_{2}(V,r)dr^2
+2 K(V,r) r^2(d\theta^2 + \sin^2\theta d\Phi^2).
\end{align}
The general gauge transformation for this perturbed metric becomes
\begin{align}
h_{\mu \nu}^{(+)} \to h_{\mu \nu}^{(+)} + \nabla_{\mu}\xi_{\nu}  + \nabla_{\nu}\xi_{\mu},
\end{align}
with
\begin{align}
\xi_\nu dx^\mu = \xi_V(V,r) dV + \xi_r(V,r)dr.
\end{align}
Under this gauge transformation, the components of the perturbed metric change as
\begin{align}
H_0 &\to H_0 - \frac{r_0}{r^2}(\xi_V + f \xi_r) + 2 \partial_V \xi_V,
\\
H_1 &\to H_1 + \partial_r \xi_V + \frac{r_0}{r^2}\xi_r + \partial_V \xi_r,
\\
H_2 &\to H_2 + 2 \partial_r \xi_r,
\\
K &\to K + \frac{1}{r}(\xi_V + f \xi_r).
\end{align}
If we choose the gauge with $H_2 = K = 0$, the residual gauge modes become
$\xi_V = -f \tilde{\eta}(V), \xi_r = \tilde{\eta}(V)$, where $\tilde{\eta}(V)$ is an arbitrary function of $V$,
and the components of the perturbed metric transform as $H_0 \to H_0 -2f \partial_V \tilde{\eta}$ and $H_1 \to H_1 + \partial_V \tilde{\eta}$.

\subsection{$\ell = 1$ odd-parity perturbations}
The general perturbed metric for the $\ell =1, m =0$ 
odd-parity linear perturbations in the Eddington-Finkelstein coordinates is given by
\begin{align}
h_{\mu \nu}^{(-)}dx^\mu dx^\nu
&=
4\sqrt{\pi/3} \sin\theta (\partial_{\theta}Y_{1,0}) d\Phi
(h_0(V,r) dV + h_1(V,r)dr)
\notag\\&=
-2 \sin^2\theta d\Phi (h_0(V,r)dV + h_1(V,r)dr).
\end{align}
The general gauge transformation for this perturbed metric becomes
\begin{align}
h_{\mu \nu}^{(-)} \to h_{\mu \nu}^{(-)} + \nabla_{\mu}\xi_{\nu}  + \nabla_{\nu}\xi_{\mu},
\end{align}
with
\begin{align}
\xi_\nu dx^\mu = -\sin^2\theta  \xi^{(-)}(V,r) d\Phi.
\end{align}
Under this gauge transformation, the components of the perturbed metric change as
\begin{align}
h_0 &\to h_0 +\partial_V\xi^{(-)},
\\
h_1 &\to h_1 - \frac{2}{r}\xi^{(-)} + \partial_r \xi^{(-)},
\end{align}
If we choose the gauge with $h_1 = 0$, the residual gauge modes become
$\xi^{(-)} = r^{2}\tilde{\zeta}(V)$, where $\tilde{\zeta}(V)$ is an arbitrary function of $V$,
and $h_0$ transforms as $h_0 \to h_0 +r^{2}\partial_V \tilde{\zeta}$.

\section{The Kerr metric with small parameter shifts}
\label{appendix:Kerr}
The Kerr metric has two parameters the mass $M$, the spin $a$.
Let us consider the shift of those parameters in Eq.~\eqref{metricKerrEF} as
\begin{align}
M &\to M + \alpha^2 \beta \delta M^{\rm (phys)},
\\
a &\to a + \alpha \beta \frac{\delta J^{\rm (phys)}}{M},
\end{align}
where $\alpha := a/M$ and $\beta$ are small parameters,
$\delta M^{\rm (phys)}$ 
and $\delta J^{\rm (phys)}$ are constants.
Introducing 
a coordinate transformation
\begin{align}
d\Phi \to d\Phi -  \alpha \beta \frac{\delta J^{\rm (phys)}}{M}  \frac{dr}{r^2},
\end{align}
and the gauge transformation at ${\cal O}(\alpha^2 \beta)$ as
\begin{align}
& g_{\mu \nu}^{\rm EF}[M + \alpha^2 \beta \delta M^{\rm (phys)}, 
a + \alpha \beta \delta J^{\rm (phys)}/M] 
\notag\\ &\to 
g_{\mu \nu}^{\rm EF}[M + \alpha^2 \beta \delta M^{\rm (phys)}, 
a + \alpha \beta \delta J^{\rm (phys)}/M]
 + \alpha^2 \beta (\nabla_\mu \xi_\nu + \nabla_\nu \xi_\mu)
\end{align}
with
\begin{align}
\xi_\mu &= \xi_\mu|_{\ell = 0} + \xi_\mu|_{\ell = 2},
\\
\xi_\mu|_{\ell = 0} dx^\mu &= \xi_{V}^{\ell = 0} dV + \xi_{r}^{\ell =0} dr,
\\
\xi_\mu|_{\ell = 2} dx^\mu &= \xi_{V}^{\ell = 2} Y_{2,0} dV + \xi_{r}^{\ell = 2} Y_{2,0} dr
-
 \frac{\xi_{S}^{\ell =2}}{\sqrt{6}}(\partial_\theta Y_{2,0})d\theta,
\end{align}
and
\begin{align}
\xi_{V}^{\ell = 0} &= - \frac{2 M \delta J^{\rm (phys)}}{r^2} + \left(1- \frac{2M}{r}\right)\tilde{\chi}(V),
\\
\xi_{r}^{\ell = 0} &= -\frac{2\delta J^{\rm (phys)}}{3r} - \tilde{\chi}(V),
\\
\xi_{V}^{\ell = 2} &= 0,
\\
\xi_{r}^{\ell = 2} &= \sqrt{\frac{\pi}{5}} \frac{4(3M+r)}{3r^2}\delta J^{\rm (phys)},
\\
\xi_{S}^{\ell = 2} &= -\sqrt{\frac{2\pi}{15}}\frac{2(2 M+r)}{r}\delta J^{\rm (phys)},
\end{align}
the metric becomes 
$g^{{\rm Kerr}+(\delta M,\delta J)}_{\mu \nu}$ with
\begin{align}
g^{{\rm Kerr}+(\delta M,\delta J)}_{\mu \nu} dx^\mu dx^\nu 
&= 
(g_{\mu \nu}^{\rm Sch} + \alpha h_{\mu \nu}^{(1,0)} + \alpha^2 h_{\mu \nu}^{(2,0)}) dx^\mu dx^\nu
-
\frac{4 \alpha \beta  \delta J^{\rm (phys)} \sin^2\theta}{r}
 dV d\Phi
\notag\\&
+ \alpha^2\beta \left[
\frac{2 \delta M^{(2,1)}_{\rm Kerr}}{r} 
+
2 \left(1-\frac{2M}{r}\right)\lambda^{(2,1)}_{\rm Kerr}
\right]dV^2
-2 \alpha^2 \beta \lambda^{(2,1)}_{\rm Kerr} dVdr
\notag\\&
+
\alpha^2\beta
\sqrt{\frac{\pi}{5}}Y_{2,0}
\bigg[
H_{0, \ell = 2}^{\rm Kerr} dV^2
+
2 H_{1,\ell = 2}^{\rm Kerr} dV dr
+
H_{2,\ell = 2}^{\rm Kerr} dr^2
\notag\\&
+
2K_{\ell = 2}^{\rm Kerr} r^2 (d\theta^2 + \sin^2\theta d\Phi^2)
\bigg]
+
{\cal O}(\alpha^3) + {\cal O}(\beta^2),
\label{eq:metricKerralphabeta}
\end{align}
where
\begin{align}
\delta M^{(2,1)}_{\rm Kerr} &= 
\delta M^{\rm (phys)} + \frac{4 M r - 6 M^2}{3 r^3}\delta J^{\rm (phys)}
\label{eq:appendixdeltaM21}
\\
\lambda^{(2,1)}_{\rm Kerr} &= - \frac{4 M \delta J^{\rm (phys)}}{3r^3} + \chi(V),
\label{eq:appendixchi}
\\
\chi(V) &:= \partial_V \tilde{\chi}(V),
\end{align}
and
\begin{align}
H_{0,\ell = 2}^{\rm Kerr} &= \frac{8 M(6M^2 - Mr - 3r^2)J^{\rm (phys)}}{3 r^5},
\\
H_{1,\ell = 2}^{\rm Kerr} &= \frac{8 M (3 M+2 r)\delta J^{\rm (phys)}}{3r^4},
\\
H_{2,\ell = 2}^{\rm Kerr}  &=
-\frac{16 M \delta J^{\rm (phys)}}{r^3},
\\
K_{\ell = 2}^{\rm Kerr} &=
- \frac{4M(2M+r)\delta J^{\rm (phys)}}{r^4}.
\end{align}
Here, we choose the coordinate system and the gauge so that ${\cal O}(\alpha \beta)$ and ${\cal O}(\alpha^2 \beta)$
terms take the similar form as in Sec.~\ref{sec:schperturbation} for $\ell = 0, 1$,
and the Regge-Wheeler gauge for $\ell = 2$.
We obtain the relation
\begin{align}
\delta M^{\rm (phys)} =  \delta M^{(2,1)}_{\rm Kerr} 
- 
\frac{4 M r - 6 M^2}{3 r^3}\delta J^{\rm (phys)}.
\label{eq:relationmphys}
\end{align}
This implies that ``the mass term'' $\delta M^{(2,1)}_{\rm Kerr}$ 
in the ${\cal O}(\alpha^2\beta)$ perturbations does not directly denote the variation of the mass of the black hole,
and the variation of the physical mass $\delta M^{\rm (phys)}$ 
is obtained by subtracting the effect of the spin from $\delta M^{(2,1)}_{\rm Kerr}$.

\end{document}